\documentclass[12pt]{article}

\usepackage{mycomment}
\usepackage{amsmath,amssymb,amsthm}
\usepackage{enumerate}
\usepackage{graphicx}
\usepackage[hang,raggedright]{subfigure}

\ifx\pdfoutput\undefined \newcount\pdfoutput \fi
\ifcase\pdfoutput
  \usepackage{hyperref}  
\else
  \usepackage[pdftex]{hyperref}
  \DeclareGraphicsExtensions{.pdf}
\fi
\hypersetup{
  pdftitle={Quaternionic Computing},%
  pdfauthor={Jose M. Fernandez, William A. Schneeberger},%
  pdfsubject={Quantum Computing},%
  pdfkeywords={Quantum Computing, quaternions, algebraic computational complexity},%
  bookmarks,bookmarksopen,%
  colorlinks,linkcolor={blue},citecolor={green},%
  urlcolor={red}}

\finaldocument

\newcommand{\transpose}[1]{{#1}^{\mathsf{t}}}%
\newcommand{\conjtrans}[1]{{#1}^\dag}
\newcommand{\bra}[1]{\langle#1\rvert}
\newcommand{\ket}[1]{\lvert#1\rangle}
\newcommand{\braket}[2]{\langle #1|#2\rangle}
\newcommand{\ketbra}[2]{\lvert #1\rangle\langle#2\rvert}
\newcommand{\norm}[1]{\lVert\,#1\,\rVert}
\newcommand{\modulus}[1]{\lvert#1\rvert}
\newcommand{\eqdef}{\ensuremath{\triangleq}}

\newcommand{\C}{\mathbb{C}} 
\newcommand{\R}{\mathbb{R}} 
\newcommand{\Hq}{\mathbb{H}} 

\newcommand{\class}[1]{\ensuremath{\mathrm{#1}}} %
\newcommand{\BQP}{\class{BQP}}

\DeclareMathOperator{\Tr}{Tr}
\DeclareMathOperator{\Diag}{Diag}

\newcommand{\ii}{\mathrm{i}}
\newcommand{\jj}{\mathrm{j}}
\newcommand{\kk}{\mathrm{k}}

\newtheorem{theorem}{Theorem}
\newtheorem{corollary}{Corollary}
\newtheorem{lemma}{Lemma}

\theoremstyle{definition}
\newtheorem{definition}{Definition}

\newcommand{\gate}[1]{\text{\tt #1}}
\newcommand{\matrice}[1]{\mathsf{#1}}

\newcommand{\SU}{\ensuremath{\mathrm{SU}}}
\newcommand{\SO}{\ensuremath{\mathrm{SO}}}
\newcommand{\Sp}{\ensuremath{\mathrm{Sp}}}

\newcommand{\qa}{\hat{\alpha}}
\newcommand{\qb}{\hat{\beta}}
\newcommand{\qe}{\hat{\eta}}
\newcommand{\qh}{\hat{h}}
\newcommand{\qG}{\hat{G}}
\newcommand{\qC}{\hat{C}}
\newcommand{\tensor}{{\cal T}}
\newcommand{\qtensor}{\hat{\tensor}}
\newcommand{\qrtensor}{\hat{\cal S}}

\newcommand{\Hilb}{\space{H}}
\newcommand{\Hc}{\Hilb_\C} 
\newcommand{\Hr}{\Hilb_\R} 

\DeclareMathOperator{\re}{Re}
\DeclareMathOperator{\im}{Im}
\DeclareMathOperator{\jm}{Jm}
\DeclareMathOperator{\km}{Km}

\DeclareMathOperator{\co}{Co}
\DeclareMathOperator{\we}{Wd}

\title{Quaternionic Computing}
\author{Jos\'{e} M. Fernandez, William A. Schneeberger}
\date{\today}

\begin{document}

\maketitle

\begin{abstract}
  We introduce a model of computation based on quaternions, which is
  inspired on the quantum computing model.  Pure states are vectors of
  a suitable linear space over the quaternions.  Other aspects of the
  theory are the same as in quantum computing: superposition and
  linearity of the state space, unitarity of the transformations, and
  projective measurements.  However, one notable exception is the fact
  that quaternionic circuits do not have a uniquely defined behaviour,
  unless a total ordering of evaluation of the gates is defined.
  Given such an ordering a unique unitary operator can be associated
  with the quaternionic circuit and a proper semantics of computation
  can be associated with it.
  
  The main result of this paper consists in showing that this model is
  no more powerful than quantum computing, as long as such an ordering
  of gates can be defined.  More concretely we show, that for all
  quaternionic computation using $n$ \emph{quaterbits}, the behaviour
  of the circuit for each possible gate ordering can be simulated with
  $n+1$ qubits, and this with little or no overhead in circuit size.
  The proof of this result is inspired of a new simplified and
  improved proof of the equivalence of a similar model based on real
  amplitudes to quantum computing, which states that any quantum
  computation using $n$ qubits can be simulated with $n+1$
  \emph{rebits}, and in this with no circuit size overhead.  

  Beyond this potential computational equivalence, however, we propose
  this model as a simpler framework in which to discuss the
  possibility of a quaternionic quantum mechanics or information
  theory.  In particular, it already allows us to illustrate that the
  introduction of quaternions might violate some of the ``natural''
  properties that we have come to expect from physical models.
\end{abstract}

\begin{mycomment}
TODO

MAJOR EDITS (OPTIONAL)
- Clarify the notion of Right vs. Left Hilbert space and discuss away...
- Give better examples of how quaternionic IT is different
 (entanglement for free, violation of causality)
- Point out that ``real'' Information Theory is also weird...
- Add octonions
- Justify the tensors in terms of the group of square roots of unity
  (Will's formalism)
\end{mycomment}

\section{Introduction}\label{sec:intro}

Quantum Computing represents yet another disconcerting puzzle to
Complexity Theory.  What we know today is that quantum computing
devices can efficiently solve certain problems, which, in appearance,
classical or probabilistic computers cannot solve efficiently.  Even
though we would like to believe that quantum computing violates the
strong Church-Turing thesis, the sore truth is that the known results
do not provide us a proof, only constituting, at best, ``strong
evidence'' thereof.

Yet, even though we cannot provide a strict separation between these
models, we do know certain inclusions between variations of these
computing models.  Perhaps the most natural variation from standard
Quantum Computing is that in which we change the domain of the state
vector amplitudes, and hence the domain of their allowed linear
transformations.

It was first shown that restricting ourselves to real amplitudes does
not diminish the power of quantum computing \cite{BV97}, and further,
that in fact rational amplitudes are sufficient \cite{ADH97}.  Both
these results were proven in the Quantum Turing Machine model, and the
respective proofs are quite technical.  Direct proofs of the first
result for the quantum circuit model stem from the fact that several
sets of gates universal for quantum computing have been found
\cite{Kit97,BMP+00,Shi02,GR02}, which involve only real coefficients.


In this paper, we introduce another possible variation on quantum
computing involving quaternionic amplitudes, and prove an equivalence
result that shows that no further computational should reasonably be
expected in this model.  In Section~\ref{sec:quantum}, we will start
by redefining quantum computing in an axiomatic fashion, which will
make it possible to easily generalise the model for other non-complex
Hilbert spaces.  We will redefine and review the results known for
computing on real Hilbert spaces in Section~\ref{sec:Real}, also
providing a new generic and structural proof of the equivalence of
this model to standard complex quantum computing.  We will introduce
the quaternionic computing model in Section~\ref{sec:Quater}, discuss
some of its peculiarities, and then show how the above proof can be
easily adapted to the quaternionic case.  In
Section~\ref{sec:Consider}, we discuss some of this result in terms of
computational complexity and also of the particularities of the
quaternionic model on in its possible ``physical'' interpretations.
Finally, we summarise our conclusions and propose further open
questions in Section~\ref{sec:Final}.

\section{Quantum Computing Revisited}
\label{sec:quantum}

The basic tenets of Quantum Computing, are as follows:
\begin{description}
\item[States.] The pure states describing the internal configuration of
  an $n$ qubit computing device are defined as 1-dimensional rays in a
  $2^n$-dimensional vector space over the complex numbers.  Over such
  a vector space, the usual inner-product defines the standard
  L2-norm, which in turn defines a proper Hilbert space\footnote{This
    is only true because we are considering finite dimensional
    inner-product spaces, which are trivially complete.}.  With
  respect to this norm, states are normally represented as unit
  vectors, up to an arbitrary phase factor $e^{i\theta}$, with
  $0\leq\theta<2\pi$.
\item[Measurement.] The canonical basis of this vector space is given
  special meaning, and called the \emph{computational basis}, in
  that it represents states which always give the same outcome when
  ``queried'' about their information content.  The states are usually
  labelled by $n$-bit strings ${\bf b} = b_1\ldots b_n$.  For a generic
  pure state $\ket{\Phi}$, the probabilities of measurement outcomes
  are given by the following rule
\begin{equation}
\Pr(\ket{\Phi}\mapsto ``{\bf b}") = \modulus{\braket{\Phi}{b}}^2
\label{eq:quantMeasure}
\end{equation}
  where $\ket{b}$ is some computational basis vector.
\item[Transformations.] Generally speaking, the transformations that
  are allowed are linear mappings.  In addition, in order for the
  quantities above to be proper probabilities, these transformations
  must preserve L2-norm.  The only relevant linear and L2-norm
  preserving operations are unitary
  transformations\footnote{Anti-unitary transformations also preserve
    L2-norm, but do not preserve inner-product and are not usually
    considered as legal quantum transformations.}.  These are usually
  represented in the matrix form in which the column vectors are the
  images of the canonical basis under the given transformation, listed
  in lexicographical order.
\item[Circuits.] The computational device is modelled as a circuit,
  which, without loss of generality, can be assumed to have the
  following characteristics
  \begin{itemize}
    \item The input to the circuit is any pure state.
    \item The circuit is an array of elementary universal gates.  
  \end{itemize}
  For example, we can choose the 2-qubit CNOT gate and arbitrary
  1-qubit rotations as a universal set.  Furthermore, we allow gates
  to operate on any two arbitrary wires, not necessarily
  contiguous\footnote{This is not the usual model, in which gates are
    restricted to act on contiguous wires.  However, this model is not
    more powerful than the later, since it can be simulated
    efficiently with at most a quadratic number of swap gates.}.
\item[Algorithms.] A quantum algorithm can be formally described as a
  classical Turing Machine, which given a classical string $\bf{x}$
  will generate a (classical) description for a quantum circuit.  The
  quantum computer can then produce an answer based on the result of
  measurements of the output wires of the quantum circuit.  Without
  loss of generality, we can assume that the circuit is to be
  evaluated with the ground state (the all zero computational base
  vector) as its initial state.  The algorithm is said to be
  \emph{efficient} if the corresponding TM runs in time polynomial on
  the size of the input $\bf{x}$, which in turns implies that circuit
  size is also polynomial.
\end{description}


From a purely abstract point of view, it can be inferred that the only
requirements of this model is that the state space has a linear
structure and a proper norm-inducing inner product, so that the
measurement rule is always sound.  Traditionally, quantum computing
has been described in terms of \emph{complex} Hilbert spaces, but in
principle, as we just discussed, a sound model of computation can be
defined on any other Hilbert space.  In particular,
in this paper we study models of \emph{real computing} and %
\emph{quaternionic computing}, based on the $2^n$-dimensional vector
spaces on the reals and the quaternions, respectively.

\section{Real Computing}\label{sec:Real}
\subsection{Definitions}

Intuitively, the real computing model is defined as a restricted
version of quantum computing, where all amplitudes in the state
vectors are required to be real numbers.  Conjugation is equivalent to
the identity operation and bras are simply transposed kets.  Similarly
the matrix dagger operator (${}^\dag$) can be replaced with the matrix
transpose operator ($\transpose{}$).

In this case, we must replace unitary transformations with orthonormal
transformations, as these are the only inner-product preserving
operations on this inner-product space.  One could conceive a model in
which the state vectors always have real amplitudes, but in which
arbitrary unitary transformations (on the complex Hilbert space) are
allowed, as long as the end result is still a real amplitude vector.
It is elementary to show that orthonormal transformations are the only
ones that have this property, and hence this model is as general as
can be, given the fact that we insist that the amplitudes be real.

\burp{TODO: statement above to be qualified for anti-unitary
  transformations.}

\subsubsection*{Rebits and States}

In quantum computing and quantum information theory, we define the
\emph{qubit} as the most elementary information-containing system.
Abstractly, the state of a qubit can be described by a 2-dimensional
state vector
\begin{equation}
\ket{\Phi} = \alpha \ket{0} + \beta \ket{1}, 
\text{ s.t. } \norm{\Phi}_2 = \sqrt{\modulus{\alpha}^2+\modulus{\beta}^2} = 1
\label{eq:Qubit}
\end{equation}
where $\ket{0}$ and $\ket{1}$ are the two canonical basis vectors for
such a 2-dimensional space.  Two vectors $\ket{\Phi}$ and
$\ket{\Phi'}$ are said to represent the same qubit value if they are
in the same 1-dimensional ray.  In other words,
\begin{equation}
\Phi \equiv \Phi' \iff \ket{\Phi}=e^{\ii\theta}\ket{\Phi'},
  \text{where } \theta\in [0,2\pi).
\label{eq:QubitEquiv}
\end{equation}

\begin{definition}[Rebit]
  The corresponding concept in real computing is called a \emph{rebit}.
  As in Equation~\ref{eq:Qubit}, its state can also be described by
  a 2-dimensional vector on the real Hilbert space
\begin{equation}
\ket{\Phi} = a \ket{0} + b \ket{1}, \; 
    \text{s.t. } \norm{\Phi}_2 = \sqrt{a^2+b^2} = 1
\label{eq:Rebit}
\end{equation}
In this case, the arbitrary phase factor can only be $+1$ or $-1$, and
the rebit equivalence relation which replaces
Equation~\ref{eq:QubitEquiv} is
\begin{align}
\Phi \equiv \Phi' & \iff \ket{\Phi}=e^{\ii\theta}\ket{\Phi'},%
  \text{where } \theta\in\{0,\pi\} \\
  & \iff \ket{\Phi}=\pm\ket{\Phi'}
\label{eq:RebitEquiv}
\end{align}
\end{definition}

Similarly as for qubits, single rebit states do have a nice
geometrical interpretation: they are isomorphic to the circumference,
having $\ket{0}$ and $\ket{1}$ at opposite extremes.  One way to see
this is to consider the locus of points in the Bloch sphere for which
$e^{\ii\theta}=1$, or in other words, those with no circular
polarisation.  Unfortunately, there is no such nice geometric
representation of an arbitrary $n$-qubit state, and we believe the
same is true for $n$-rebit states.

The computational basis vectors for a rebit are still $\ket{0}$ and
$\ket{1}$, and for arbitrary $n$-rebit systems they can also be
represented as $n$-bit strings.  The measurement rule in defining the
probabilities of obtaining the corresponding bit string as a result is
essentially the same as Equation~\ref{eq:quantMeasure},
\begin{equation}
\Pr(\ket{\Phi}\mapsto ``{\bf b}") = \modulus{\braket{\Phi}{b}}^2
                                  = \braket{\Phi}{b}^2 
\label{eq:realMeasure}
\end{equation}
where in this case we can drop the modulus operator $\modulus{\cdot}$,
because it is redundant.

One physical interpretation that can be given for rebits or rebit
systems is that of a system of photons, where we use the polarisation
in the usual manner to carry the information.  However, these photons
are restricted to having zero circular polarisation, and being
operated upon with propagators which never introduce circular
polarisation, i.e.\ orthonormal operators.  The computational basis
measurements are still simple polarisation measurements in the
vertical-horizontal basis.

\burp{LONG-TERM TODO: Mention here about thing getting weird,
  concerning ``real entanglement.''}

\subsubsection*{Real Circuits and Real Computational Complexity}

We can also define and construct real circuits, as a restriction of
quantum circuits.  Topologically, they are the same, as we will still
require them to be constructed only with reversible gates.  Since
orthonormal matrices, like unitary matrices, are preserved under the
tensor algebra that describes circuit constructions (see
\cite{BFH02,BFH04} for more details on this formalism), it is
sufficient to require that the elementary gates be orthonormal.  With
this, we are assured that the overall circuit transformation will be
norm-preserving.  We can then define a measurement rule for circuit
states, which will yield classical results with probabilities exactly
as in Equation~\ref{eq:realMeasure}.  As was noted before, this rule
is completely general and does not depend on the field on which the
inner-product space of states is defined.

\subsubsection*{Real Algorithms}
To complete the definition of this computational model, we must define
what it means for such real computing devices to ``compute'' or to
``solve a problem.''  For that, we simply restrict the definition of
a quantum algorithm given above.

\begin{definition}[Real Algorithm]\label{def:realAlg}
  A \emph{real algorithm} is defined as a classical TM, which on
  (classical) input $\bf{x}$ will generate a (classical) description
  of a \emph{rebit circuit}.  The result of measurement of the final
  state $\ket{\Phi}$ of the rebit circuit is post-processed by the TM
  to produce its final (classical) answer.
\end{definition}

The TM can be viewed as having access to a universal circuit evaluator
or oracle, which will produce a classical answer ${\bf b}$, with the
probabilities defined in Equation~\ref{eq:realMeasure}.  It is important
to note that no matter what classical post-processing the classical
Turing Machine does after obtaining an answer from the Oracle, its
final answer ultimately only depends on the outcome probabilities.  In
other words, from the TM's point of view, it does not matter if the
circuit is physically constructed or just simulated by the Oracle, nor
does it matter what technology was used or what mathematical
abstraction was employed in its simulation.  What matters is that the
outcome probabilities of the Oracle be the same as those of circuit
description provided by the TM.
                                
\subsection{Previously Known Results}
\label{sec:prevKnown}

From a Complexity Theory point of view, the first question that arises
naturally is how does this real computing model compare with the
quantum computing one.  In other words, can the problems which are
efficiently solved by a quantum algorithm also be solved by an
efficient real algorithm?

For the Quantum Turing Machine model, the answer was previously known
to be ``Yes''.  Even though, it is not explicitly stated as such, the
following theorem is traditionally attributed to Bernstein and
Vazirani, as it can be easily deduced from the results in \cite{BV97}.

\begin{theorem}[Bernstein, Vazirani]\label{thm:realQTM}
  Any Quantum Turing Machine can be approximated sufficiently well by
  another, whose transition matrix only contains computable real
  numbers of the form $\pm\cos(k R)$ and $\sin(kR)$, where $k$
  is an integer and
$$R=\sum_{i=1}^\infty 1/2^{2^i}.$$
\end{theorem}

The need for having such transcendental amplitudes was eventually
removed.  By using transcendental number theory techniques, Adleman,
Demarrais, and Huang showed in~\cite{ADH97}, that, in fact, only a few
rational amplitudes were required, in particular only the set $\{0,\pm
1,\pm 3/5,\pm 4/5\}$.


It is important to note that Theorem~\ref{thm:realQTM} does not apply
directly to circuits, or at least not in a completely trivial manner.
The constructions in the proof are relatively elaborate and rely
heavily on techniques of Turing Machine engineering.  Nonetheless,
quantum circuits were shown to be equivalent to Quantum Turing
Machines by A.C.-C.~Yao in \cite{Yao93}.  In principle, the construction
of that proof could be used to show that quantum circuits do not
require states with complex amplitudes to achieve the same power as
any complex-valued circuit or QTM.  

However, the celebrated universality result of Barenco, Bennett,
Cleve, DiVicenzo, Margolus, Shor, Sleator, Smolin, and Weinfurter
\cite{BBC+95} provides a first step towards a proof of that fact, as
they show that \gate{CNOT} and arbitrary 1-qubit gates form a
universal set of gates for quantum circuits.  While arbitrary 1-qubit
gates can contain complex amplitude transitions, more recent results
have produced ever smaller sets of universal gates, which are
comprised only of real amplitude transitions.  The following is just a
sample list of such results:
\begin{itemize}
\item \gate{TOFFOLI}, \gate{HADAMARD}, and $\pi/4$-rotation, by
  Kitaev \cite{Kit97} in 1997.
\item \gate{CNOT}, \gate{HADAMARD}, $\pi/8$-rotation by Boykin,
  Mor, Pulver, Roychowdhury, and Vatan \cite{BMP+00} in 2000.
\item \gate{TOFFOLI} and \gate{HADAMARD}, by Shi \cite{Shi02} in
  2002, with a simpler proof by Aharonov \cite{Aha03} in 2003.
\item Controlled $\theta$-rotations, by Rudolph and Grover \cite{GR02}
  in 2002
\end{itemize}

The motivation behind these results was to come up with the simplest
possible gates, given the fact that quantum states in nature can and
will have arbitrary complex amplitudes, and thus, so will their
unitary propagators.  The fact that the simpler sets involve only real
numbers was \emph{a priori} just a ``desirable side-effect.''  Our
motivation, however, is completely different.  We play a different
game: suppose that all we had were these mysterious ``rebits,'' unable
to enter complex amplitudes.  What could we do then?  Because of this
motivation, our proof will have a different flavour.  In fact, the
proof is completely general in that it works with any universal set of
gates.  In particular, it will work with gates which have arbitrary
complex transition amplitudes.  In other words, in proving the
following, more general theorem, we will completely ignore the above
results.  That will allow us to recycle its proof later on in
Section~\ref{sec:Quater}.


\begin{theorem}\label{thm:RebitSim}
  Any $n$-qubit quantum circuit constructed with gates of degree $d$
  or less (possibly including non-standard complex coefficients gates)
  can be exactly simulated with an $n+1$ rebit circuit with the same
  number of gates of degree at most $d+1$.
\end{theorem}

\subsection{A New Proof of Equivalence}
\label{sec:newProof}

\subsubsection{The Underlying Group Theory}

The idea behind the proof is to make use of the fact that the group
$\SU(N)$ can be embedded into the group $\SO(2N)$.  We provide an explicit
embedding $h$.\footnote{Independently, Aharonov \cite{Aha03} has also
  used this mapping recently to provide a simple proof that
  \gate{TOFFOLI} and \gate{HADAMARD} are universal.}  While this
mapping is not unique, what is special about it is that it has all the
necessary properties for us to define a sound simulation algorithm
based on it.  This mapping is defined as follows.  Given an arbitrary
unitary transformation $U$, its image $O=h(U)$ is
\begin{equation}
U \stackrel{h}{\mapsto} O = h(U) \eqdef
    \left(\begin{array}{r|r} \re(U)&\im(U)\\ \hline -\im(U)&\re(U)\end{array}\right)
    \label{eq:RealMapBlock}
\end{equation}
where the $\re$ and $\im$ operators return the real and imaginary
parts of a complex number, respectively, and applied to complex
matrices, return the matrix composed of the real and imaginary parts
of each entry.  Note also, that if we define the following formal
tensor
\begin{equation}
    \tensor \eqdef \left(\begin{array}{rr} \re & \im \\ -\im & \re \end{array} \right)
    \label{eq:RealTensor} 
\end{equation}
we can express the definition of $h$ more simply as
\begin{equation}
U \stackrel{h}{\mapsto} O = h(U) = \tensor \otimes U
    \label{eq:RealMapTensor}
\end{equation}
  
The first fundamental property that this mapping must have for us to
use it effectively in a simulation is the following.

\begin{theorem}
  \label{thm:iso}
  Let $G_N$ represent the image of\/ $\SU(N)$ under $h$.  Then $h$ is a
  proper group isomorphism between $\SU(N)$ and $G_N$, and $G_N$ is a
  subgroup of\/ $\SO(N)$.
\end{theorem}

\begin{proof}
  It is easy to see that any matrix in $G_N$, which will have the form
  of Equation~\ref{eq:RealMapBlock}, will have a unique inverse
  image, and hence that $h$ is an injective mapping.  The following
  lemma is sufficient to show that $h$ is a group homomorphism.

\begin{lemma}\label{lem:hProjector}
  Let $A$ and $B$ be any two arbitrary $N\times N$ matrices, then
  $h(AB)=h(A)h(B)$.
\end{lemma}

\begin{proof}
  The first step is to obtain a simple matrix multiplication rule for
  matrices, using the operators $\re$ and $\im$.  For arbitrary complex
  numbers $\alpha$ and $\beta$, we have that
  \begin{eqnarray}
    \re(\alpha\beta)& = & \re(\alpha)\re(\beta)-\im(\alpha)\im(\beta) \nonumber \\
    \im(\alpha\beta)& = & \re(\alpha)\im(\beta)+\im(\alpha)\re(\beta) 
    \label{eq:multRuleComplex}
  \end{eqnarray}
  Since these rules hold for the products of all of their entries, it
  is then easy to see that this same multiplication rule will also
  hold for complex matrices.  In other words, we can substitute
  $\alpha$ and $\beta$ in Equation~\ref{eq:multRuleComplex} with any
  two arbitrary complex matrices $A$ and $B$ which are multipliable,
  to get
  \begin{eqnarray}
    \re(AB)& = & \re(A)\re(B)-\im(A)\im(B) \nonumber \\
    \im(AB)& = & \re(A)\im(B)+\im(A)\re(B) 
    \label{eq:multRuleComplexMatrix}
  \end{eqnarray}
  We are now equipped to verify our claim
  \begin{align}
    h(A)h(B)
    & = (\tensor\otimes A)(\tensor\otimes B) \nonumber \\
    & = \left(\begin{array}{r|r}\re(A)&\im(A)\\\hline -\im(A)&\re(A)\end{array}\right)
         \left(\begin{array}{r|r}\re(B)&\im(B)\\\hline -\im(B)&\re(B)\end{array}\right)\nonumber\\
    & = \left(\begin{array}{r|r}\re(A)\re(B)-\im(A)\im(B) & \re(A)\im(B)+\im(A)\re(B) \\
        \hline -\im(A)\re(B)-\re(A)\im(B) & -\im(A)\im(B)+\re(A)\re(B)\end{array}\right)\nonumber\\
    & = \left(\begin{array}{r|r}\re(AB)&\im(AB)\\\hline -\im(AB)&\re(AB)\end{array}\right)\nonumber\\
    & = \tensor\otimes AB = h(AB) \label{eq:TensorMult}
  \end{align}
\end{proof}

Finally, we want to show that $G_N\subset\SO(2N)$.  This is equivalent
to showing that all the images $O=h(U)$ are orthonormal, i.e.\ that
$\transpose{O}=O^{-1}$.  Since by Lemma~\ref{lem:hProjector} $h$ is a
group homomorphism, it maps inverse elements into inverse elements,
i.e.~$h(U^{-1})=h(U)^{-1}$.  Since $U$ is unitary, we have that
\begin{equation}
  O^{-1}=h(U)^{-1}=h(U^{-1})=h(U^\dag)
  \label{eq:Oinverse}
\end{equation}
while the following lemma will give us an expression for
$\transpose{O}$.

\begin{lemma}\label{lem:hHermitian}
  Let $A$ be an arbitrary $N\times N$ complex matrix, then
  $h(A^\dag)=\transpose{h(A)}$.
\end{lemma}
\begin{proof}
  By definition of $h$ and by transposition rules of block matrices,
  we have
  \begin{align}
    \transpose{h(A)} & = \transpose{(\tensor\otimes A)} \nonumber  \\
&= \transpose{\left(\begin{array}{r|r}
      \re(A)&\im(A)\\ \hline
      -\im(A)&\re(A)\end{array}\right)} \nonumber \\
&= \left(\begin{array}{r|r}
    \transpose{\re(A)}&\transpose{-\im(A)}\\ \hline
    \transpose{\im(A)}&\transpose{\re(A)} \end{array}\right) \nonumber \\
&= \left(\begin{array}{r|r}
    \re(A^\dag)&\im(A^\dag)\\ \hline
    -\im(A^\dag)&\re(A^\dag)\end{array}\right)\nonumber \\
&= \tensor\otimes A^\dag = h(A^\dag)
    \label{eq:transposeTensor}
\end{align}
where we also used the following generic matrix identities
\begin{align}
    \re(A^\dag) & =  \transpose{\re(A)} \nonumber \\
    \im(A^\dag) & = -\transpose{\im(A)} \label{eq:matrixIdentities}.
\end{align}
\end{proof}

In particular, we have that
$\transpose{O}=\transpose{h(U)}=h(U^\dag)=h(U^{-1})=O^{-1}$, and we
are done proving Theorem~\ref{thm:iso}.
\end{proof}  

The fact that $h$ is a group isomorphism is important, because it
implies that $G_N$ is preserved under ``serial'' circuit construction.
In other words, it means that if we have real circuits that simulate
the quantum circuits with operators $U$ and $V$, then we can simulate
a quantum circuit with operator $UV$ by simply putting both real
circuits together.  This suggests a way in which to decompose the
problem of simulating a generic quantum circuit, i.e.\ by constructing
the real circuit one level at a time.

\subsubsection{The Simulation Algorithm}

Let $C$ be a generic $n$-qubit quantum circuit with operator $U_C$,
composed of $s$ elementary gates.  The simulation algorithm will
consist of the following steps:
\begin{enumerate}[{\bf Step~1.}]
\item Serialise the given circuit by finding an ordering of its gates,
  so that they can be evaluated in that order, one by one.  In other
  words, find a total order of the circuit gates, such that 
  $U_C = U^{(s)} U^{(s-1)} \dots U^{(2)} U^{(1)}$.
  \label{item:TO}
\item For each gate $g\in\{1,\dots,s\}$ in the above ordering, replace
  the $n$-ary operation $U^{(g)}$, corresponding to the $g$-th gate,
  with an adequate real circuit $O^{(g)}$ simulating it.
  \label{item:map}
\item Construct the overall real circuit $C'$ by concatenating the
  circuits for each level $g$, in the same order as defined in
  Step~\ref{item:TO}.  This is, if $O_C$ is the operator for $C'$,
  then let $O_C = O^{(s)}\dots O^{(2)}O^{(1)}$.
  \label{item:circuit}
\item Write a description of the real circuit $C'$ and of its input
  state and ask the real computing ``oracle'' to provide the result of
  a measurement on its final state.
  \label{item:I&M}
\item Perform the classical post-processing on the result of the
  measurement and provide a classical answer.\label{item:post}
\end{enumerate}
The algorithm, as described so far, is not completely defined.  In what
follows, we will derive, one by one, the missing details.

\begin{figure}[ht]
  \centering
  \includegraphics{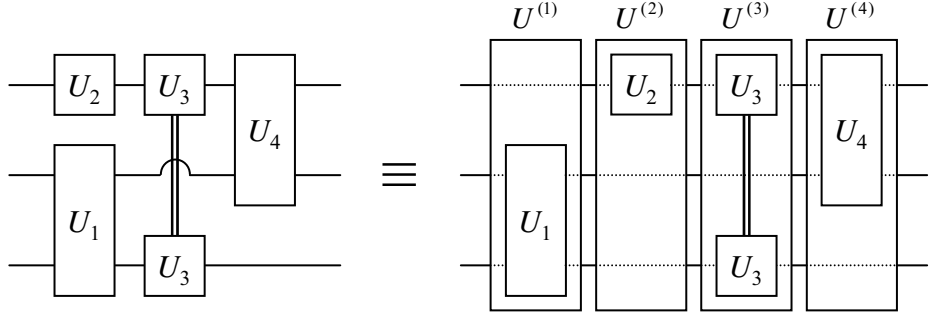}
  \caption[Serialisation of a Quantum Circuit]
  {Serialisation of the quantum circuit $C$ in Step~\ref{item:TO}.}
  \label{fig:serialize}
\end{figure}

First, the total order in Step~\ref{item:TO} can be obtained by doing
a topological sort of the circuit's directed graph.  This can be done
efficiently in time polynomial in the size of the
circuit\footnote{These orderings, because of the fact that they can be
  found efficiently, are the base of the ``strong'' equivalence of the
  circuit and the Turing Machine models of computation.}.  The effects
of Step~\ref{item:TO} on $C$ are depicted in
Figure~\ref{fig:serialize}.

\subsubsection{Constructing the Real Circuit}

In principle, each of the elementary quantum gates $g$ is described by
a unitary operator defined on the $d$-qubit complex Hilbert space.  We
can assume without loss of generality that these gates are described
in the input to the simulation algorithm as $2^d\times 2^d$
matrices\footnote{In fact, what we are given are finite-precision
  \emph{approximations} of these matrices.}, which we denote with
subscripted capitals.  Thus, the $g$-th gate has associated to it a
$d$-ary gate operator $U_g$ (with typically $d=1,2$).

However, in the context of a circuit the operator fully describing the
action of gate $g$ is an $N$-ary operator acting on all $n$ qubits,
which depends not only on $U_g$ but also on the positions of the wires
on which $g$ acts.  We denote this operators with superscripted
capitals.  Thus, after serialisation of the circuit $C$ in
Step~\ref{item:TO}, these operators $U^{(g)}$ will correspond to the
$g$-th level of the serialised version of $C$.

\begin{figure}[h]
  \centering
  \includegraphics{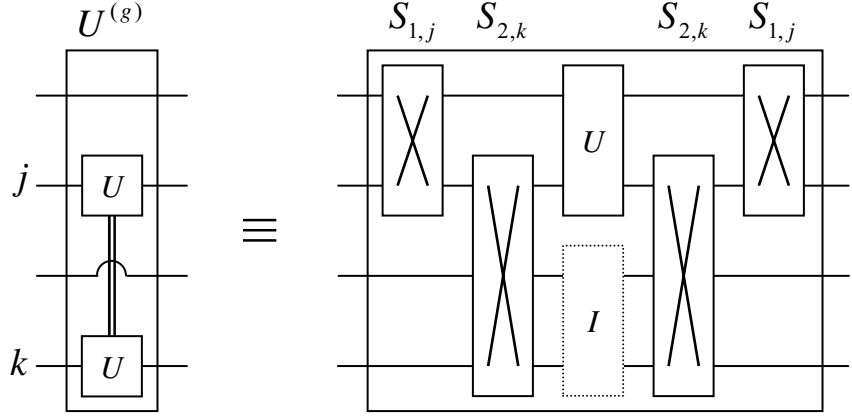}
  \caption[Obtaining a Formula for an $N$-ary Quantum Circuit]
  {Obtaining an expression for the $N$-ary circuit operator $U^{(g)}$.}
  \label{fig:paddingU}
\end{figure}

In general, the $g$-th gate will be a $d$-ary gate operating on wires
with indices $j_1<j_2<\dots<j_d$, not necessarily contiguous, with the
associated circuit operator $U^{(g)}$, which will depend on
$j_1,\dots,j_d$.  For example, in the case of a 2-qubit gate $g$
operating on the $j$-th and $k$-th wires, $1\leq j<k\leq n$, $U^{(g)}$
can be expressed in terms of its elementary gate $U_g$ as follows
\begin{eqnarray}
    U^{(g)}(j,k) 
        & =      & S_{1,j}\,S_{2,k} \, (U_g\otimes I_{n-2}) \, S_{2,k}\,S_{1,j} \nonumber\\
        & \eqdef & S_g \, (U_g\otimes I_{n-2}) \, S_g
    \label{eq:paddingDefTwo} 
\end{eqnarray}
where $I_m$ is the identity operator for $m$ qubits, $S_{i,j}$ is the
$n$-qubit swap operator acting on wires $i$ and $j$, and $S_g\eqdef
S_{j,k}$ is a shorthand for describing the necessary swap operator for
the $g$-th gate.  The logic behind Equation~\ref{eq:paddingDefTwo}
is explained graphically in Figure~\ref{fig:paddingU}.  Note however,
that this conversion using swap gates is not itself part of the
simulation, but only a mathematical convenience to be used later.
These swaps gates will not be included in the final real circuit $C'$
and do not represent a computational overhead.

As for Step~\ref{item:map}, the isomorphism $h$ readily suggests a
method for substituting each of the $s$ levels of the original quantum
circuit $C$.  Let $\Hc^d$ be the $2^d$-dimensional complex Hilbert
space on which $U_g$ acts, and let $\Hr^d$ be the
$2^{d+1}$-dimensional real Hilbert space on which its image
$O_g=h(U_g)$ acts.  If $g$ is a $d$-qubit gate, then $O_g$ operates on
$d+1$ rebits.  We thus have an extra wire, and it is not \emph{a
  priori} clear how to map the original $d$ quantum wires with these
$d+1$ real wires.  To resolve this ambiguity, we need to define how we
associate the base vectors of $\Hc^d$ with those of $\Hr^d$.  

We use the columns of the tensor $\tensor$ defining $h$ in
Equation~\ref{eq:RealMapTensor}, to define the following mappings
between $\Hc^d$ and $\Hr^d$.  Let $\ket{\Phi}$ be an arbitrary state
vector in $\Hc^d$, and let $\tensor=[\tensor_0|\tensor_1]$,
\begin{align}
\ket{\Phi}
  &\stackrel{h_0}{\longmapsto}\ket{\Phi_0}
     \eqdef\tensor_0\otimes\ket{\Phi} =
     \left(\begin{array}{rr}\re\\-\im\end{array}\right)\otimes\ket{\Phi}
  \label{eq:rVectorMap}\\
  &\stackrel{h_1}{\longmapsto}\ket{\Phi_1}
     \eqdef\tensor_1\otimes\ket{\Phi} =
     \left(\begin{array}{rr}\im\\\re\end{array}\right)\otimes\ket{\Phi}
  \label{eq:iVectorMap}
\end{align}
Note that the images $\ket{\Phi_0}$ and $\ket{\Phi_1}$ are mutually
orthogonal in $\Hr^d$.  In addition, both $h_0$ and $h_1$ are proper
linear homomorphisms, as can be easily verified given the
distributivity of the tensor product with matrix addition.

The base vectors $\ket{b}$ of $\Hc^d$ are column vectors with all zero
entries, except with a 1 at the integer value $j$ of $\mathbf{b}$;
i.e.~$\braket{j}{b}=1$, and $\braket{k}{b}=0, k\neq j$.  Thus, it is
easy to see what these basis vectors are mapped to:
\begin{eqnarray}
\ket{b} & \mapsto & \ket{b_0} = \ket{0} \otimes \ket{b}\\
        & \mapsto& \ket{b_1} = \ket{1} \otimes \ket{b}
\label{eq:BasisMap}
\end{eqnarray}

\begin{figure}[h]
  \centering
  \includegraphics{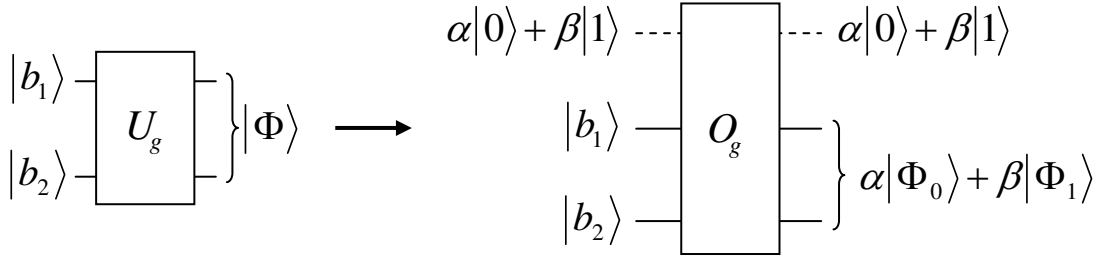}
  \caption[Simulation of a 2-qubit Quantum Gate]%
  {Simulation of an individual elementary binary quantum gate, by a
    tertiary real gate.  Note that in general, for non classical
    inputs, the final state of $O_g$ cannot be factored like in the
    example shown.}
  \label{fig:simulGate}
\end{figure}

\burp{TODO: Fix notation such that $b_0$ and $b_1$ are replaced with
  $b_\R$ and $b_\C$ or something like that to avoid confusion with
  Figure~\ref{fig:simulGate}...}

These homomorphisms define the semantics to give to each of the $d+1$
real wires on which $O_g$ acts, as is shown in
Figure~\ref{fig:simulGate}.  When the original quantum gate takes
$\ket{b}$ as input, the corresponding real gate $O_g$ has two possible
base vectors $\ket{b_0}$ or $\ket{b_1}$ as inputs.  This corresponds
to having an extra wire at the top of the gate with value $\ket{0}$ or
$\ket{1}$ respectively, and the base state $\ket{b}$ in the bottom $d$
wires.  Finally, note that since $U_g$ is represented as a matrix of
constant dimension, then $O_g$ is also a small matrix, which can be
computed from $U_g$ and written down in constant time.

Even though we have defined how to simulate ``out-of-context'' $d$-ary
elementary quantum gates, we have not yet explained how to simulate
them in their corresponding positions in the circuit $C$.  In other
words, we still have to describe how to simulate the $N$-ary operators
$U^{(g)}$.  Again, the isomorphism $h$ comes to the rescue: we will
simulate $U^{(g)}$ by finding an $(n+1)$-rebit circuit that computes its
image $O^{(g)}=h(U^{(g)})$ under~$h$.  Unfortunately, we cannot simply
construct this circuit from the matrix definition of $h(U^{(g)})$,
because it is a huge matrix and that would require exponential time.
However, $U^{(g)}$ is a very simple $N$-ary operator: it is after all
just a $d$-ary gate, which has a succinct description given by
Equation~\ref{eq:paddingDefTwo}.  Since it involves at most $d$
qubits, then the circuit $O^{(g)}$ only needs to involve those same
wires and one other extra rebit.

\begin{figure}[h]
  \centering
  \includegraphics[scale=0.85]{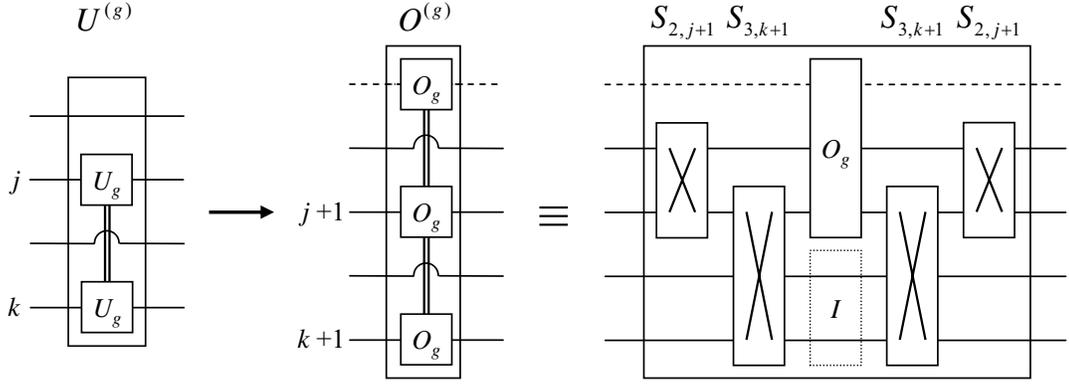}
  \caption[Obtaining a Formula for an $(N+1)$-ary Real Circuit]
  {Obtaining an expression for the $(N+1)$-ary circuit $O^{(g)}$.}
  \label{fig:paddingO}
\end{figure}

At this point, we have to make a further apparently arbitrary choice,
i.e.\ which one of the $n-d$ other available wires will play the role
of the ``top'' rebit for the $O_g$ gate?  In other words, where shall
we place the extra wire required for implementing $O^{(g)}$?  The
answer comes from the homomorphisms $h_0$ and $h_1$ in Equations
(\ref{eq:rVectorMap}) and~(\ref{eq:iVectorMap}), respectively.  They
are also automatically defined on the state space $\Hc^N$ of the whole
circuit, and hence they generate the same wire semantics as for
isolated $d$-ary gates: the extra wire must be at the top of the
circuit, as is shown in Figure~\ref{fig:paddingO}.  Similarly, as in
Equation~\ref{eq:paddingDefTwo}, we have for the case where $d=2$,
an expression for $O^{(g)}$ in terms of $O_g$.
\begin{eqnarray}
    O^{(g)}(j,k) &=& S_{2,j+1}\,S_{3,k+1}\,
                     (O_g\otimes I_{n-2})\,
                     S_{3,k+1}\,S_{2,j+1} \nonumber \\
            &\eqdef& S'_g \, (O_g\otimes I_{n-2}) \, S'_g
    \label{eq:paddingDefTwoReal}
\end{eqnarray}
where again we define $S'_g$ for convenience, and $j$ and $k$ are the
indices of the wires on which gate $g$ acts on the original circuit
$C$.

\begin{figure}[h]
  \centering
  \includegraphics{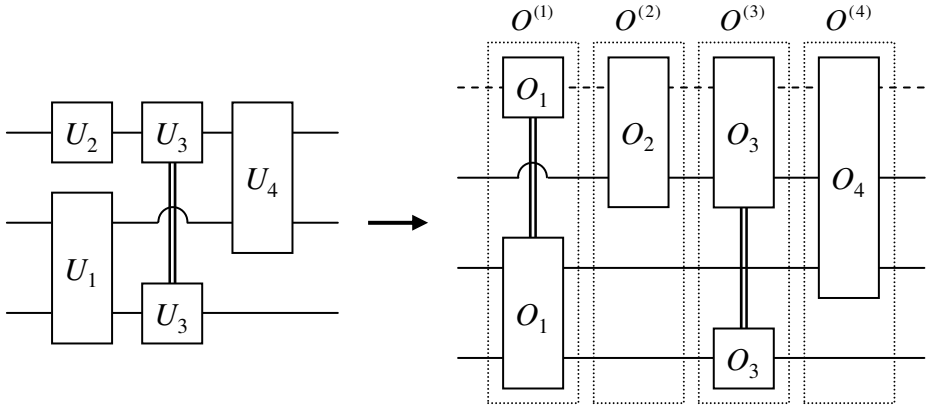}
\caption[Simulation of a Quantum Circuit by a Real Circuit]
   {Simulation of a quantum circuit by a real circuit.}
\label{fig:realCircuit}
\end{figure}

We now have a simple and well defined scheme for constructing the
desired simulating circuit $C'$.  In Step~\ref{item:circuit}, we will
construct $C'$ by concatenating the real circuits for the $N$-ary
operators $O^{(g)}$.  One important characteristic of this scheme is
that we are \emph{reusing} the extra wire needed for each gate, each
time using the same top wire.  This is illustrated in
Figure~\ref{fig:realCircuit}.  Even though they act on the whole space
$\Hc^N$, the $O^{(g)}$ operators are simply $(d+1)$-ary gates put in
context, and they can be described in a succinct manner requiring only
a constant number of symbols.  Therefore, the overall size of the
description for $C'$ will be linear in the size of the initial
description of $C$ which was given as input.

What is remarkable about this scheme, is that despite its simplicity,
it gives precisely what we wanted, this is, that the final operator
$O_C$ be in some sense as similar as possible to the operator $U_C$ of
the original circuit.  In fact, we have the following third nice
property of our simulation.

\begin{lemma}\label{lem:circularity}
  The inverse image of $O_C$ is precisely $U_C$, i.e.~$O_C=h(U_C)$.
\end{lemma}

\begin{proof}
  Because of the serialisation of Step~\ref{item:TO}, we have that
  $U_C = U^{(s)}\dots U^{(2)}U^{(1)}$.  We use this and the group
  isomorphism properties of $h$ from Lemma~\ref{lem:hProjector} to obtain the
  following expression for its image
  \begin{align}
    h(U_C)
       &= h(U^{(s)}\dots U^{(1)}) \nonumber\\
       &= h(U^{(s)})\dots h(U^{(1)}) \nonumber\\
       &= \prod_{\substack{i\,=\,0,\\g\,=s-i}}^{s-1} h(U^{(g)})\nonumber
  \intertext{We can now use the expression of
    Equation~\ref{eq:paddingDefTwo} to substitute for $U^{(g)}$,}
       &= \prod \; h(\,S_g\,(U_g\otimes I_{n-2})\,S_g) \nonumber\\
       &= \prod \; h(S_g) \; h(U_g\otimes I_{n-2}) \; h(S_g)\nonumber
  \intertext{Since $S_g$ is composed only of 0's and 1's, we have that $\re(S_g)=S_g$ 
    and $\im(S_g)=0$.  Furthermore, we have that $S'_g = I_1\otimes S_g$ 
    from their definition in Equations (\ref{eq:paddingDefTwo}) 
    and~(\ref{eq:paddingDefTwoReal}), and thus,}
       &= \prod \; (I_1 \otimes S_g) \; h(U_g\otimes I_{n-2}) \; (I_1 \otimes S_g)\nonumber\\
       &= \prod \; S'_g \:  h(U_g\otimes I_{n-2}) \; S'_g \nonumber
  \intertext{However, the tensor product is just a formal operation, and its
    associativity property holds even with a  tensor of operators 
    like $\tensor$.  Hence, we have}
       &= \prod \; S'_g \:  [ \tensor \otimes (U_g\otimes I_{n-2})\, ]\; S'_g \nonumber\\
       &= \prod \; S'_g \:  [\, (\tensor \otimes U_g)\otimes I_{n-2}\, ]\; S'_g \nonumber\\
       &= \prod \; S'_g \:  [h(U_g) \otimes I_{n-2}\,]\; S'_g \nonumber\\
       &= \prod \; S'_g \:  (O_g \otimes I_{n-2}) \; S'_g\nonumber
  \intertext{which with the padding expression of $O^{(g)}$ in 
    Equation~\ref{eq:paddingDefTwoReal} finally gives}
       &= \prod_{\substack{i\,=\,0,\\g\,=s-i}}^{s-1} \; O^{(g)} = O_C\:.
  \end{align}
\end{proof}

\subsubsection{Circuit Initialisation and Measurement}

Having described how to construct the real circuit $C'$ from the
original circuit $C$, we still have to address the issue of how to
initialise $C'$ in Step~\ref{item:I&M}, and furthermore of how to
interpret and use its measurements to simulate the initial quantum
algorithm in Step~\ref{item:post}.

Let $\ket{\Psi}$ represent the initial state given to $C$, and let
$\ket{\Phi}$ be its image under $U_C$, i.e.\ the final state of the
circuit before measurement.  If we think back of the two homomorphisms
$h_0$ and $h_1$ from $\Hc^N$ to $\Hc^N$, induced by $h$, we have two
logical choices for initialising the corresponding real circuit $O_C$,
the states $\ket{\Psi_0}$ and $\ket{\Psi_1}$.  Which should we choose,
and in either case what will the output look like?  The answer to the
latter question is given by the following lemma.

\begin{lemma}
  \label{lem:initChoice}
  The images of $\ket{\Psi_0}$ and $\ket{\Psi_1}$ in the real circuit
  $C'$ are
  \begin{eqnarray}
    O_C\ket{\Psi_0} &= \tensor_0\otimes\ket{\Phi} = \ket{\Phi_0} \\
    O_C\ket{\Psi_1} &= \tensor_1\otimes\ket{\Phi} = \ket{\Phi_1} 
    \label{eq:homoMap1Real}
  \end{eqnarray}
\end{lemma}

\begin{proof}
  As in the proof of Lemma~\ref{lem:hProjector}, all we require are the
  matrix multiplication rules of
  Equation~\ref{eq:multRuleComplexMatrix}
  \begin{align}
    O_C\ket{\Psi_0}
      &= (\tensor\otimes U_C)(\tensor_0\otimes\ket{\Psi_0}) \nonumber \\
      &= \left(\begin{array}{r|r}\re(U_C)&\im(U_C)\\ \hline -\im(U_C)&\re(U_C)
                 \end{array}\right)
         \left(\begin{array}{r}\re(\ket{\Psi_0})\\\hline\im(\ket{\Psi_0})
                 \end{array}\right)\nonumber\\
      &=  \left(\begin{array}{r}\re(U_C)\re(\ket{\Psi_0})+\im(U_C)\im(\ket{\Psi_0})\\\hline
                              -\im(U_C)\re(\ket{\Psi_0})+\re(U_C)\im(\ket{\Psi_0})
                 \end{array}\right) \nonumber\\
      &= \left(\begin{array}{r}\re(U_C\ket{\Psi_0})\\\hline -\im(U_C\ket{\Psi_0})
                 \end{array}\right) \nonumber\\
      &= \tensor_0\otimes(U_C\ket{\Psi_0}) \nonumber \\
      &= \tensor_0\otimes\ket{\Phi} = \ket{\Phi_0}
\intertext{With the same method, we can obtain a similar expression for $\Phi_1$, i.e.}
    O_C\ket{\Psi_1}
      &= (\tensor\otimes U_C)(\tensor_1\otimes\ket{\Psi_1}) \nonumber \\
      &= \left(\begin{array}{r|r}\re(U_C)&\im(U_C)\\ \hline -\im(U_C)&\re(U_C)
                 \end{array}\right)
         \left(\begin{array}{r}\im(\ket{\Psi_1})\\\hline\re(\ket{\Psi_1})
                 \end{array}\right) = \ldots \nonumber\\
      &= \tensor_1\otimes\ket{\Phi} = \ket{\Phi_1}
  \end{align}
\end{proof}

Let us assume for a moment ---and in fact, this is without loss of
generality--- that the original circuit was to be initialised with
some base vector $\ket{x}$, with a final state $\ket{\Phi}=U\ket{x}$.
Again, there are two possible choices for initialising the
corresponding real circuit, namely $\ket{x_0}=\ket{0}\ket{x}$ and
$\ket{x_1}=\ket{1}\ket{x}$.  What would then be the output of the
simulated circuit in either case?  In the very special case that
$\ket{\Phi}$ is also a base vector, then we would have
$\ket{\Phi_0}=\ket{0}\ket{\Phi}$ and $\ket{\Phi_1}=\ket{1}\ket{\Phi}$,
and thus, in either case, the bottom $n$-wires would contain the right
answer and we can ignore the top wire.  But when $\ket{\Phi}$ is some
arbitrary pure state, neither purely real nor purely imaginary, we
cannot give such a nice semantic to the top wire.  In particular, it
might be entangled with the rest of the wires, and hence we cannot
factor the final state.

Nonetheless, what is surprising is that if we \emph{trace out} the top
wire, in all cases we will get the same statistics and furthermore
that we will obtain the right statistics, i.e.\ the same as if we had
used the original quantum circuit $C$\@.  More formally, we have

\begin{lemma}\label{lem:measure}
  Let $\ket{\Phi}$ be an arbitrary $n$-qubit pure state, and let
  $\rho_0=\Tr_1\ketbra{\Phi_0}{\Phi_0}$ and
  $\rho_1=\Tr_1\ketbra{\Phi_1}{\Phi_1}$ represent the partial traces
  obtained by tracing out (i.e.\ forgetting about) the top wire.  Then
  we have that
  \begin{align}
  \rho_0 &= \rho_1, \\
  \Diag\,(\rho_0) &= \Diag\,(\rho_1) = \Diag\,(\ketbra{\Phi}{\Phi}).
  \end{align}
\end{lemma}

\begin{proof}
  The partial trace of the first wire of an arbitrary density operator
  given in block matrix form
  $$\rho = \left(\begin{array}{c|c}A & B\\ \hline C & D\end{array}\right) $$
  is given by,
  \begin{eqnarray}
    \Tr_1 (\rho) & = & [I_n|0]\,\rho\,[I_n|0]^\dag + [0|I_n]\,\rho\,[0|I_n]^\dag  \nonumber \\
                 & = & A + D
    \label{eq:traceOut}
  \end{eqnarray}
  In particular, we have that
  \begin{align}
    \ketbra{\Phi_0}{\Phi_0} & = (\tensor_0\otimes\ket{\Phi})\;
                     \transpose{(\tensor_0\otimes\ket{\Phi})}\nonumber
\intertext{which by applying transposition rules for block matrices and 
Equation~\ref{eq:multRuleComplexMatrix} gives}
    &= \left(\begin{array}{r}\re(\ket{\Phi})\\\hline-\im(\ket{\Phi})\end{array}\right)\;
       \left(\begin{array}{r|r}\re(\bra{\Phi})&      \im(\bra{\Phi})\end{array}\right)\nonumber\\
    &= \left(\begin{array}{r|r}
              \re(\ket{\Phi})\re(\bra{\Phi})&\re(\ket{\Phi})\im(\bra{\Phi})\\\hline
             -\im(\ket{\Phi})\re(\bra{\Phi})&-\im(\ket{\Phi})\im(\bra{\Phi})\end{array}\right)
         \label{eq:densityOp0}
\intertext{and similarly for $\ket{\Phi_1}$,}
    \ketbra{\Phi_1}{\Phi_1} & = (\tensor_1\otimes\ket{\Phi})\;
                     \transpose{(\tensor_1\otimes\bra{\Phi})}\nonumber\\
    &= \left(\begin{array}{c|c}
              -\im(\ket{\Phi})\im(\bra{\Phi})&\im(\ket{\Phi})\re(\bra{\Phi})\\\hline
             -\re(\ket{\Phi})\im(\bra{\Phi})&\re(\ket{\Phi})\re(\bra{\Phi})\end{array}\right)
         \label{eq:densityOp1}
\intertext{By symmetry, we thus have the same expression for both partial traces}
  \rho_0=\rho_1& =\re(\ket{\Phi})\re(\bra{\Phi}) - \im(\ket{\Phi})\im(\bra{\Phi}) \nonumber\\
               & =\re\,(\ketbra{\Phi}{\Phi})
  \end{align}
  Since $\ketbra{\Phi}{\Phi}$ is hermitian, its diagonal entries are
  all real, and therefore it has the same diagonal entries as $\rho_0$
  and $\rho_1$.
\end{proof}

In other words, combining this with Lemma~\ref{lem:initChoice}, we
arrive to the conclusion that it does not matter what we set as the
initial value of the top wire, $\ket{0}$ or $\ket{1}$.  Furthermore,
it is easy to verify that any 1-rebit state will do, whether pure or
even totally mixed, as long as it is unentangled and uncorrelated with
the bottom wires.

\subsection{Further Considerations and Consequences}

\subsubsection{Complexity of simulation}


In general, if we initially have a $d$-qubit gate, the new gate will
be a $(d+1)$-rebit gate.  However, if $U_g$ contains only real entries,
then $O_g = I \otimes U_g$, which means that in this particular case
the top rebit need not be involved, and therefore the new gate is the
same as the original.  If the whole quantum circuit we are given is
constructed with such real gates, then we are in luck and we do not
require the extra rebit at all.  In the general complex case, however,
the circuit width is at most one more than that of the original
circuit.


However, one non-negligible consequence of our simulation is that any
parallelism that the original circuit may have had is lost after we
serialise the circuit in Step~\ref{item:TO} of the simulation
algorithm.  While it might be still possible to parallelise parts of
the real circuit $C'$ (e.g.\ where we had real gates in the $C$), in
the worst case, if all gates in $C$ require complex amplitudes, then
the top wire is always used and the circuit depth for $C'$ is equal to
its gate count $s$.  This is a consequence of our decision to reuse
the same wire as the ``top wire'' for each gate.  However, it is
possible to reduce this depth increase at the cost of using several
``top wires'' and re-combining them towards the end of the circuit.
This will result in only a $O(\log s)$ increase in circuit depth.


Finally, as we have mentioned before, the overall classical pre- and
post-processing requires little computational effort.  Converting a
description for the original circuit $C$ into $C'$ requires time
linear in the size of the circuit description, i.e.~$O(s)$.
Post-processing will be exactly the same as for the original quantum
algorithm, since the statistics of measuring the bottom wires of $C'$
(or any subset thereof) will be exactly the same as those of measuring
the wires of $C$, as per Lemma~\ref{lem:measure}.

\subsubsection{Universality} 

We knew already, from the previous results mentioned in
Section~\ref{sec:prevKnown}, that it is possible to express any
quantum circuit in terms of real gates only.  If we had not known
already that fact, we could have presumed that quantum circuits would
be described and given to us in terms some universal set of gates
containing at least one non-real, complex gate.  In that case,
Theorem~\ref{thm:RebitSim} would provide a proof that a real universal
set could be constructed, simply by replacing any non-real gates by
its image under $h$.

One advantage of this technique is that it does this conversion with
very limited overhead in terms of width, requiring 1 extra rebit
\emph{for the whole circuit}, and not an extra rebit for every
substituted gate, as might have been expected.  In addition to its
usefulness in Section~\ref{sec:Quater}, this is one of the reason that
we believe that this particular version of the equivalence theorem is
interesting of its own, when compared to previously known results.  In
particular, the fact that it provides a much tighter bound on
simulation resources needed, might prove useful in the study of lower
quantum complexity classes and possibly in quantum information theory.

\subsubsection{Interpretation}


With Lemma~\ref{lem:measure}, we are left with a curious paradox:
while we require an extra rebit to perform the simulation, we do not
care about its initial or its final value.  In particular, it can be
anything, even the maximally mixed state.  So, what is this rebit
doing?

Let $\Hilb_0$ and $\Hilb_1$ be the orthogonal subspaces, each of
dimension $N$, spanned by the $\ket{b_0}$ and $\ket{b_1}$ base vectors
of Equations \ref{eq:rVectorMap} and~\ref{eq:iVectorMap},
respectively.  If a state $\ket{\Phi}$ has only real amplitudes then
$\ket{\Phi_0}\in\Hilb_0$ and $\ket{\Phi_1}\in\Hilb_1$.  For a generic
$\ket{\Phi}$, however, $\ket{\Phi_0}$ and $\ket{\Phi_1}$ are not
contained in either subspace, but in the space spanned by both, i.e.\
the complete rebit space $\Hr^{n}$.  In that case, the top rebit
will not be just $\ket{0}$ or $\ket{1}$ but some superposition
thereof.

In other words, it somehow keeps track of the phase (angle) of the
representation of $\ket{\Phi}$ in rebit space with respect to these
subspaces.  The \gate{CNOT} gate (or any other real gate) does not
change this phase factor.  However, as arbitrary gates with complex
transition amplitudes affect this phase factor, their effect is
simulated by ``recording'' this change in the top rebit.  How we
initialise the top rebit gives an arbitrary initial phase to the
representation of $\ket{\Phi}$, but as we saw, this initial phase does
not affect statistics of the bottom wires, and thus can be set to any
value.  However, how this phase has been changed by previous complex
gates will affect the bottom rebits in subsequent complex gates, in a
similar fashion as the \emph{phase kickback} phenomenon in many
quantum algorithms\footnote{With the noticeable difference that phase
  kickback would not work if the top qubit were maximally mixed...}.
That is why that top rebit is needed.

\begin{mycomment}
Conjecture1: Top bit always disentangled from the rest of the rebits??
FALSE!!!
Conjecture2: Top rebit can be totally mixed??
YES, I sort of checked...
\end{mycomment}

\section{Quaternionic Computing}\label{sec:Quater}

This section closely mimics Section~\ref{sec:Real}.  First we define
what we mean by quaternionic computing, making sure that it is a
sensible model.  We then prove an equivalence theorem with quantum
computing, by using the same techniques as those of
Theorem~\ref{thm:RebitSim}.

\subsection{Definitions}
\label{sec:quaterDefs}

\subsubsection{Quaternions}

Quaternions were invented by the Irish mathematician William Rowan
Hamilton in 1843, as a generalisation of complex numbers.  They form a
non-commutative, associative division algebra.  A quaternion is
defined as
\begin{equation}
\qa = a_0 + a_1\ii + a_2\jj + a_3\kk\label{eq:quaterDef}
\end{equation}
where the coefficients $a$ are real numbers and $\ii$, $\jj$, and
$\kk$ obey the equations
\begin{equation}
\ii\ii = \jj\jj = \kk\kk = \ii\jj\kk = -1\label{eq:quaterUnitRoot}
\end{equation}

Multiplication of quaternions is defined by formally multiplying two
expressions from Equation~\ref{eq:quaterDef}, and recombining the
cross terms by using Equation~\ref{eq:quaterUnitRoot}.  It is very
important to note that while all non-zero quaternions have
multiplicative inverses they are \emph{not commutative}
\footnote{While the square roots of $-1$ are anti-commutative,
  e.g.~$\ii\jj=-\jj\ii$, this is not true in general,
  i.e.~\mbox{$\qa\qb\neq -\qb\qa$}.}.  Thus, they form what is called
a \emph{division algebra}, sometimes also called a \emph{skew field}.

The quaternion conjugation operation is defined as follows:
\begin{equation}
  \qa^\star = a_0 - a_1\ii - a_2\jj - a_3\kk
  \label{eq:quaterConj}
\end{equation}
where for clarity, we represent with the (non-standard) symbol
$(^\star)$ in order to distinguish it from complex conjugation
represented with $(^*)$.  With this conjugation rule, we can define the
modulus of a quaternion as
\begin{equation}
  \modulus{\qa} = \sqrt{\qa\qa^\star} = \sqrt{a_0^2+a_1^2+a_2^2+a_3^2}
  \label{eq:qmodulus}
\end{equation}
Furthermore, the usual vector inner product has the required
properties (i.e.\ it is norm defining), and a proper Hilbert space can
be defined on any quaternionic linear space.

It is also possible to \emph{complexify} the quaternions, this is, to
represent them in terms of complex numbers only.  Let $\qa$ be an
arbitrary quaternion, then we define its \emph{complex} and
\emph{weird} parts as
\begin{eqnarray}
\co(\qa)&\eqdef&a_0+a_1\ii\\
\we(\qa)&\eqdef&a_2+a_3\ii
  \label{eq:WeirdComplex}.
\end{eqnarray}
We can then decompose $\qa$ in its complex and weird part as follows:
\begin{eqnarray}
\qa &=&a_0+a_1\ii+a_2\jj+a_3\kk \nonumber\\
    &=&(a_0+a_1\ii)+(a_2+a_3\ii)\jj \nonumber \\
    &=&\co(\qa)+\we(\qa)\jj
    \label{eq:complexify}
\end{eqnarray}
This equation allows us to derive multiplication rules, similar to
those of Equation~\ref{eq:multRuleComplex}
\begin{eqnarray}
\co(\qa\qb)& = & \co(\qa)\co(\qb)-\we(\qa)\we^*(\qb) \nonumber \\
\we(\qa\qb)& = & \co(\qa)\we(\qb)+\we(\qa)\co^*(\qb) \label{eq:multRuleQuater}
\end{eqnarray}
where we define $\co^*(\qa)\eqdef[\co(\qa)]^*$, and similarly for the
weird part $\we^*(\qa)\eqdef[\we(\qa)]^*$.  It is interesting to note
how the non-commutativity of quaternions is made apparent by the fact
that neither identity in Equation~\ref{eq:multRuleQuater} is
symmetric with respect to $\qa$ and $\qb$, unlike their equivalent for
complex numbers (Equation~\ref{eq:multRuleComplex}), because in
general $\co^*(\qa)\neq\co(\qa)$ and $\we^*(\qa)\neq\we(\qa)$.  We can
also rewrite Equation~\ref{eq:qmodulus} for the modulus as 
\begin{equation}
  \modulus{\qa} = \sqrt{\modulus{\co(\qa)}^2 + \modulus{\we(\qa)}^2}
  \label{eq:qmodulusCoWd}
\end{equation}
which is very similar to the modulus definition for complex numbers.
Finally, we have the following useful identities
\begin{align}
  \co(\qa^\star) &= \hspace{1em}\co^*(\qa)\nonumber\\ \we(\qa^\star) &= -\we(\qa)
  \label{eq:coweIdent}
\end{align}

\subsubsection{Quaterbits}

Similarly as in quantum information theory, we can define the
quaternionic equivalent to the qubit, as the most elementary
quaternionic information system, the \emph{quaterbit}.\ \footnote{The
  name ``quits'' has also been suggested \cite{Pou02} and
  abandoned...}

\begin{definition}[Quaterbit]
  A \emph{quaterbit} is a 2-level system with quaternionic amplitudes.
  It can be represented by a unit vector $\ket{\Phi}$ in a
  2-dimensional quaternionic Hilbert space, i.e.
  \begin{equation}
    \ket{\Phi} = \qa \ket{0} + \qb \ket{1}, \;
          \text{s.t. } \norm{\Phi}_2 = \sqrt{\modulus{\qa}^2+\modulus{\qb}^2}
    \label{eq:Quaterbit}
  \end{equation}
  up to an arbitrary \emph{quaternionic} phase factor.  Indeed, we have that
  \begin{equation}
    \Phi \equiv \Phi' \iff \ket{\Phi}=\qe \ket{\Phi'},%
    \text{where } \modulus{\qe}=1.
    \label{eq:QuaterbitEquiv}
  \end{equation}
\end{definition}

The canonical values of the quaterbit correspond to the canonical
basis $\ket{0}$ and $\ket{1}$ of that vector space, and are given the
same semantics just as before.  Similarly, we can define $n$-quaterbit
states, with the same canonical basis as for rebits and qubits.  With
this definition, the measurement rule in Equation~\ref{eq:quantMeasure}
is still sound and we adopt it axiomatically.

Quaternions are often used in computer graphics to represent rotations
of the 3D Euclidean space.  However, contrary to rebits or qubits, we
have not found a nice geometric interpretation for the state space of
even a single quaterbit.


\subsubsection{Quaternionic Circuits}
\label{sec:quater}

For the sake of clarity, let us distinguish the conjugate transpose
operation for quaternion and complex matrices by representing them
differently with the $(^\ddag)$ and $(^\dag)$ symbols, respectively.
As before, the only relevant linear transformations $Q$ that preserve
$l_2$ norm on this vector space are the \emph{quaternionic unitary
  transformations}, which have the same property $Q^\ddag=Q^{-1}$ as
complex unitary transformations.  They form the so-called
\emph{symplectic} group which is represented as $\Sp(N)$.

\burp{What about anti-unitary quaternionic transformations???}

Thus armed with linear, inner-product preserving operations, we can in
principle define \emph{quaternionic circuits} in a similar fashion as
we defined quantum and real circuits.  Unfortunately, we cannot apply
the same definition of computation semantics as before, and thus
cannot define \emph{quaternionic algorithms} in the same way either.
The reason is simple and quite surprising: the output of a
quaternionic circuit is not uniquely defined!

To see this, consider the following property of the matrix tensor
product, i.e.~the distributivity of the tensor product with the
regular matrix product.
\begin{equation}
  \label{eq:tensorDistrib}
  \matrice{
    (A\otimes B)\cdot(C\otimes D)=(A\cdot C)\otimes(B\cdot D)
  }
\end{equation}
where $\matrice{A,B,C,D}$ are arbitrary matrices.  This equation is in
general true for any \emph{commutative} semiring and for
non-commutative semirings only if $\matrice{C}$ and $\matrice{D}$ are
0-1 matrices.

\begin{figure}[h]
  \centering
  \includegraphics[scale=0.85]{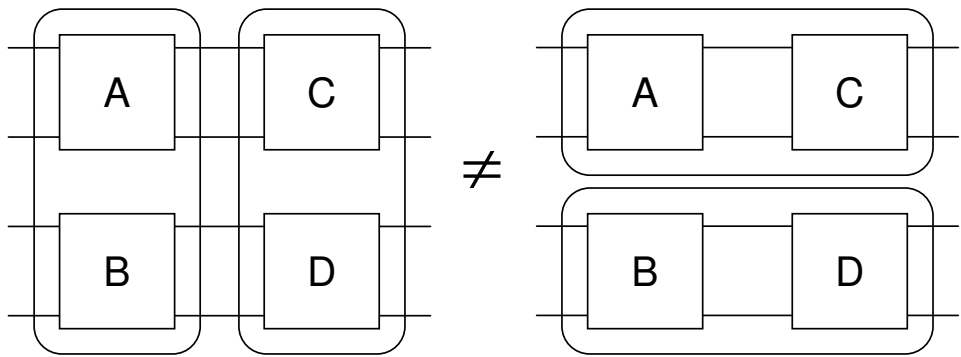}
\caption[Effects of Quaternionic Non-Commutativity on Quaternionic Circuits]
{Effects of quaternionic non-commutativity on quaternionic circuits.
  The operator for the circuit on the left is obtained by combining
  the operators ``vertically'', by taking the tensor product first;
  this corresponds to the operator on the left side of
  Equation~\ref{eq:tensorDistrib}.  The operator for the right circuit
  is obtained by combining them ``horizontally'' first, and gives the
  expression on the right hand side of the same equation.}
\label{fig:quaterWeird}
\end{figure}

Suppose now that the matrices $\matrice{A,B,C,D}$ correspond to the
gate transformations in the circuit depicted in
Figure~\ref{fig:quaterWeird}.  Then, the fact that
Equation~\ref{eq:tensorDistrib} does not hold means that the two
different ways shown there of combining the gates will yield different
operators for the circuit.  Furthermore, even if we initialise in both
cases with the same input, we will obtain different output statistics.

To further illustrate this paradox it is useful to think of the states
of a circuit in terms of \emph{temporal cuts} in the circuit graph
(see \cite{Fer04} for a more detailed description of this formalism).
We can think of the set of all possible states of a given circuit
graph as its discrete ``space-time continuum.''  The circuit topology
defines an ordering on this set that is naturally associated with a
state being ``before'' or ``after'' another.  It is however only
\emph{partially ordered} as some states are temporally incomparable,
i.e.~those corresponding to cuts of the graph that cross each other.
Each topological sort of the circuit graph is one of the many possible
total orderings of the set of cuts, or in other words a \emph{chain}
in the poset (partially-ordered set) of cuts, also corresponding, as
we saw in Section~\ref{sec:newProof}, to an evaluation sequence of
gates.  In more physical terms, each of these chains or total orders
corresponds to a possible path in the space-time continuum of the
circuit.

When Equation~\ref{eq:tensorDistrib} holds, we are guaranteed that the
overall operator over each and all of these paths \emph{will be the
  same}.  However, in the case of the quaternionic circuits, we can
expect each of these paths to give a different answer.  Which of these
many paths (for a poly-size circuit, there are exponentially many of
them) is the ``correct'' one?  Which one is somehow privileged by
nature?  Which one should we choose to be the ``computational output''
of the circuit?  The fact is that we do not know how to resolve this
ambiguity, and without doing it, it is not completely clear what
``the'' model of quaternionic computing should consist of.  

We can get out of this impasse by allowing for a ``parametrised''
notion of a quaternionic algorithm.  

\begin{definition}[Output of a Quaternionic Circuit]
  Let $\qC$ be a quaternionic circuit of size $s$ and let
  $\sigma=(\sigma_1,\dots,\sigma_s)$ represent one of the possible
  topological sorts of the corresponding circuit graph.  We denote by
  $Q_\sigma$ the operator of the circuit $\qC$ under $\sigma$, which
  is obtained when the gates are combined one-by-one following the
  ordering in $\sigma$, i.e.~
  $$Q_\sigma = Q^{(s)} Q^{(s-1)} \cdots Q^{(2)} Q^{(1)},$$
  where
  $Q^{(i)}$ is the (in-context) operator corresponding to the $i$-th
  gate in $\sigma$.
\end{definition}

\begin{definition}[Quaternionic Algorithm]\label{def:quaterAlg}
  A \emph{quaternionic algorithm} is defined as a classical TM, which
  on (classical) input $\bf{x}$ will generate a (classical)
  description of a quaternionic circuit $\qC$ \emph{and} a (classical)
  description of one of its possible topological sorts $\sigma$.  The
  result of measurement of the final state
  $\ket{\Phi}=Q_\sigma\ket{\Psi_0}$, where $\ket{\Psi_0}$ is the
  default initial state, is then post-processed by the TM to produce
  its final (classical) answer.
\end{definition}

Relative to this somewhat unsatisfying notion of quaternionic
computation, we are still able to obtain the following equivalence
result.  This theorem is the main result of this article, and its
proof is very heavily inspired from that of
Theorem~\ref{thm:RebitSim}.

\begin{theorem}\label{thm:QuaterSim}
  Let $\qC$ be any $n$-quaterbit circuit of size $s$, composed of
  gates of degree at most $d$ and let $\sigma$ be \emph{any}
  topological sort of $\qC$.  Then, there exists a quantum circuit of
  $n+1$ qubits, employing the same number of gates, each of degree at
  most $d+1$, that \emph{exactly} simulates the operator $Q_\sigma$.
\end{theorem}

\subsection{Proof of Main Theorem}

\subsubsection{More Group Theory}

As before, the proof is based on the (lesser known) fact that $\Sp(N)$
can be embedded into $\SU(2N)$.  We provide a mapping from one to the
other, which is very similar to the one from $\SU(N)$ to $\SO(2N)$.

The mapping $\qh$ from $\Sp(N)$ to $\SU(2N)$ is defined similarly to
the one from $\SU(N)$ to $\SO(2N)$ given in
Equation~\ref{eq:RealMapBlock}
\begin{align}
Q \stackrel{\qh}{\mapsto} U=\qh(Q) & \eqdef 
    \left(\begin{array}{r|r} \co(Q)&\we(Q)\\ \hline -\we^*(Q)&\co^*(Q)\end{array}\right)
    \label{eq:QuaterMapBlock}
\intertext{or equivalently in its tensor form, as in Equation~\ref{eq:RealTensor}}
    & = \left(\begin{array}{cc} \co & \we \\ -\we^* & \co^* \end{array} \right)\otimes Q \nonumber \\
    & \eqdef \qtensor\otimes Q
    \label{eq:QuaterMapTensor}
\end{align}

At this point, what we need to show is that this $\qh$ is also a group
isomorphism, in other words the equivalent of Theorem~\ref{thm:iso}.

\begin{theorem}
  \label{thm:isoQuaternion}
  Let $\qG_N$ represent the image of $\Sp(N)$ under $\qh$.  Then $\qh$
  is a proper group isomorphism between $\Sp(N)$ and $\qG_N$, and
  $\qG_N$ is a subgroup of $\SU(N)$.
\end{theorem}

Thanks to the tensor formalism, we do not need to construct the proof
in full detail, as we did for Theorem~\ref{thm:iso}.  The only thing
we need to show are equivalent statements to those of Lemmas
\ref{lem:hProjector} and~\ref{lem:hHermitian}.

\begin{lemma}\label{lem:qhProjector}
  Let $A$ and $B$ be any two arbitrary $N\times N$ quaternion
  matrices, then $\qh(AB)=\qh(A)\qh(B)$.
\end{lemma}

\begin{proof}
  As before, it is simple to verify that the quaternion multiplication
  rules in Equation~\ref{eq:multRuleQuater} also generalise to any
  multipliable quaternionic matrices $A$ and $B$.  Thus we have that
\begin{align}
  \qh(A)\qh(B)
&= (\qtensor\otimes A)(\qtensor\otimes B) \nonumber \\
&= \left(\begin{array}{r|r} \co(A)&\we(A)\\ \hline -\we^*(A)&\co^*(A)  \end{array}\right)
   \left(\begin{array}{r|r} \co(B)&\we(B)\\ \hline -\we^*(B)&\co^*(B)\end{array}\right) \nonumber \\
&= \left(\begin{array}{r|r} \co(A)\co(B)-\we(A)\we^*(B) & \co(A)\we(B)+\we(A)\co^*(B) \\
                      \hline -\we^*(A)\co(B)-\co^*(A)\we^*(B) & -\we^*(A)\we(B)+\co^*(A)\co^*(B)\end{array}\right)%
                  \nonumber \\
&= \left(\begin{array}{rr} \co & \we \\ -\we^* & \co^* \end{array}\right)\otimes AB=\qtensor\otimes AB=\qh(AB)
  \label{eq:qhProjector}
\end{align}
\end{proof}

\begin{lemma}\label{lem:qhHermitian}
  Let $A$ be an arbitrary $N\times N$ quaternion matrix, then
  $\qh(A^\ddag)=\conjtrans{\qh(A)}$.
\end{lemma}

\begin{proof}
  Similarly to the proof of Lemma~\ref{lem:hHermitian}, we require the
  following matrix identities, which are easily verified
  \begin{align}
    \co(A^\ddag)  &=\hspace{1em}\co(A)^\dag \nonumber \\
    \co^*(A^\ddag)&=\hspace{1em}\co^*(A)^\dag \nonumber \\
    \we(A^\ddag)  &=-\we^*(A)^\dag \label{eq:QuaterMatrixIdentities}. 
  \end{align}
  We then have that
  \begin{eqnarray}
    \qh(A)^\dag & = &
    \left(\begin{array}{r|r} \co(A)&\we(A)\\ \hline -\we^*(A)&\co^*(A)\end{array}\right)^\dag \nonumber \\
&=& \left(\begin{array}{r|r} \co(A)^\dag&-\we^*(A)^\dag\\ \hline \we(A)^\dag&\co^*(A)^\dag\end{array}\right) \nonumber \\
&=& \left(\begin{array}{r|r} \co(A^\ddag)&\we(A^\ddag)\\ \hline -\we^*(A^\ddag)&\co^*(A^\ddag)\end{array}\right) \nonumber \\
&=& \qtensor\otimes A^\ddag = h(A^\ddag)
\end{eqnarray}
\end{proof}

\subsubsection{The Simulation Algorithm}
Let $\qC$ be a quaternionic circuit composed of $s$ elementary gates
of at most $d$ quaterbits, let $\sigma$ represent a path in its
space-time continuum (i.e.~one of its possible total ordering or
equivalently a topological sort of the circuit graph), and let
$Q_\sigma$ be the corresponding quaternionic linear operator.  Then
the quantum simulation algorithm for $\qC$ under $\sigma$ will be very
similar to that described in Section~\ref{sec:newProof}.

\begin{enumerate}[\bf{Step} 1]
\item Serialise the given circuit $\qC$ according to $\sigma$,
  i.e.~such that $Q_\sigma = Q^{(s)} Q^{(s-1)} \dots Q^{(2)} Q^{(1)}$.
  \label{item:qTO}
\item For each gate $g\in\{1,\dots,s\}$ in the ordering defined by
  $\sigma$, let $g_1<\dots<g_d$ be the wires on which the $d$-ary gate
  $Q_g$ acts.  Replace $Q^{(g)}$ with $U^{(g)}$ the appropriately
  padded $(n+1)$-qubit operator for the quantum gate $U_g=\qh(Q_g)$
  acting on wires $g_1+1<\dots<g_d+1$ and the top qubit wire.
  \label{item:qmap}
\item Construct the overall quantum circuit $C$ by concatenating the
  circuits for each level $g$, in the same order as defined in
  Step~\ref{item:qTO}.  That is, if $U$ is the operator for $C$, then
  let $U = U^{(s)}\dots U^{(2)}U^{(1)}$.
  \label{item:qcircuit}
\item Write a description of the quantum circuit $C$ and of its
  (classical) input state and ask the quantum computing ``oracle'' to
  provide the result of a measurement on its final state.
  \label{item:qI&M}
\item Perform exactly the same classical post-processing on the result
  as the original quaternionic algorithm.\label{item:qpost}
\end{enumerate}

The construction of the circuit as described in
Section~\ref{sec:newProof} is purely formal, and does not depend at
all on the actual gates and operators.  In particular, other than
circuit operator algebra, the proof of Lemma~\ref{lem:circularity}
only required that $h$ be a group isomorphism, fact which we have
already established for $\qh$.  Thus we can claim the following
equivalent lemma.

\begin{lemma}\label{lem:qcircularity}
  The inverse image of $U$ is precisely $Q$, i.e.~$U=\qh(Q)$.
\end{lemma}

\subsubsection{Initialisation and Measurement}

We can maintain the same semantics for $\ket{\Phi_0}$ and
$\ket{\Phi_1}$, such as defined in Equations~(\ref{eq:rVectorMap})
and~(\ref{eq:iVectorMap}), by the using the columns $\qtensor_0$ and
$\qtensor_1$ of the new tensor $\qtensor=[\qtensor_0|\qtensor_1]$,

\begin{align}
\ket{\Phi}
  &\stackrel{\qh_0}{\mapsto}\ket{\Phi_0}
     \eqdef\qtensor_0\otimes\ket{\Phi} =
     \left(\begin{array}{c}\co\\-\we^*\end{array}\right)\otimes\ket{\Phi}
  \label{eq:q0VectorMap}\\
  &\stackrel{\qh_1}{\mapsto}\ket{\Phi_1}
     \eqdef\qtensor_1\otimes\ket{\Phi} =
     \left(\begin{array}{c}\we\\\co^*\end{array}\right)\otimes\ket{\Phi}
  \label{eq:q1VectorMap}
\end{align}

With these definitions, we have the same base cases for setting the
top wire, thanks to the following lemma, equivalent to
Lemma~\ref{lem:initChoice}.

\begin{lemma}
  \label{lem:qInitchoice}
  Let $\ket{\Psi}$ be any $n$-quaterbit state, then we have that the
  images of $\ket{\Psi_0}$ and $\ket{\Psi_1}$ in the quantum circuit
  $C$ are
  \begin{eqnarray}
    U\ket{\Psi_0} &= \qtensor_0\otimes\ket{\Phi} = \ket{\Phi_0} \\
    U\ket{\Psi_1} &= \qtensor_1\otimes\ket{\Phi} = \ket{\Phi_1} 
    \label{eq:homoMap1Quater}
  \end{eqnarray}
\end{lemma}

\begin{proof}
  With the quaternion matrix multiplication rules obtained from
  Equation~\ref{eq:multRuleQuater}, we have
  \begin{align}
    U\ket{\Psi_0}
      &= (\tensor\otimes Q)(\tensor_0\otimes\ket{\Psi_0}) \nonumber \\
      &= \left(\begin{array}{r|r}\co(Q)&\we(Q)\\ \hline -\we^*(Q)&\co^*(Q)
                 \end{array}\right)
         \left(\begin{array}{r}\co(\ket{\Psi_0})\\\hline-\we^*(\ket{\Psi_0})
                 \end{array}\right)\nonumber\\
      &=  \left(\begin{array}{r}\co(Q)\co(\ket{\Psi_0})-\we(Q)\we^*(\ket{\Psi_0})\\\hline
                              -\we^*(Q)\co(\ket{\Psi_0})-\co^*(Q)\we^*(\ket{\Psi_0})
                 \end{array}\right) \nonumber\\
      &= \left(\begin{array}{r}\co(Q\ket{\Psi_0})\\ \hline -\we^*(Q\ket{\Psi_0})
                 \end{array}\right) \nonumber\\
     &= \tensor_0\otimes(Q\ket{\Psi_0}) \nonumber \\
     &= \tensor_0\otimes\ket{\Phi} = \ket{\Phi_0}
\intertext{And similarly for $\Phi_1$, i.e.}
    U\ket{\Psi_1}
     &= (\tensor\otimes Q)(\tensor_1\otimes\ket{\Psi_1}) \nonumber \\
     &= \left(\begin{array}{r|r}\co(Q)&\we(Q)\\ \hline -\we^*(Q)&\co^*(Q)
                \end{array}\right)
        \left(\begin{array}{r}\we(\ket{\Psi_1})\\\hline\co^*(\ket{\Psi_1})
                \end{array}\right) = \ldots \nonumber\\
     &= \tensor_1\otimes\ket{\Phi} = \ket{\Phi_1}
  \end{align}
\end{proof}

Finally, we need to show that as before we can initialise with any
qubit value in the top wire, ignore it at measurement, and still get
the same statistics as we would have with the original quaternionic
circuit.  For that, we have to show that the equivalent of
Lemma~\ref{lem:measure} is still true.

\begin{lemma}\label{lem:qMeasure}
  Let $\ket{\Phi}$ be an arbitrary $n$-quaterbit state, $\ket{\Phi_0}$
  and $\ket{\Phi_1}$ its images under $\qh_0$ and $\qh_1$, and
  $\rho_0$ and $\rho_1$ be their respective partial traces when the
  first qubit wire is traced out.  Then,
  \begin{equation}
    \Diag\,(\rho_0) = \Diag\,({\rho_1}) = \Diag\,(\ketbra{\Phi}{\Phi}).
    \label{eq:qmeasure}
  \end{equation}
\end{lemma}


\begin{proof}
  The expressions for the non-reduced density operators are given by
  \begin{align}
    \ketbra{\Phi_0}{\Phi_0} 
    & = (\qtensor_0\otimes\ket{\Phi})\;(\qtensor_0\otimes\bra{\Phi})^\dag
           \nonumber\\
    &= \left(\begin{array}{r}\co(\ket{\Phi})\\\hline-\we^*(\ket{\Phi})\end{array}\right)\;\
       \left(\begin{array}{r|r}\co(\bra{\Phi})&\we(\bra{\Phi})\end{array}\right)\nonumber\\
    &= \left(\begin{array}{r|r}
        \co(\ket{\Phi})\co(\bra{\Phi})&\co(\ket{\Phi})\we(\bra{\Phi})\\\hline
       -\we^*(\ket{\Phi})\co(\bra{\Phi})&-\we^*(\ket{\Phi})\we(\bra{\Phi})\end{array}\right)
         \label{eq:qdensityOp0}
\intertext{and similarly,}
    \ketbra{\Phi_1}{\Phi_1} 
    & = (\qtensor_1\otimes\ket{\Phi})\;(\qtensor_1\otimes\bra{\Phi})^\dag
           \nonumber\\
    &= \left(\begin{array}{r}\we(\ket{\Phi})\\\hline\co^*(\ket{\Phi})\end{array}\right)\;\
       \left(\begin{array}{r|r}-\we^*(\bra{\Phi})&\co^*(\bra{\Phi})\end{array}\right)\nonumber\\
    &= \left(\begin{array}{r|r}
       -\we(\ket{\Phi})\we^*(\bra{\Phi})&\we(\ket{\Phi})\co^*(\bra{\Phi})\\\hline
       -\co^*(\ket{\Phi})\we^*(\bra{\Phi})&\co^*(\ket{\Phi})\co^*(\bra{\Phi})\end{array}\right)
         \label{eq:qdensityOp1}
  \end{align}
  As before, the reduced density operators are the sum of block
  matrices in the diagonal, which, unlike in Lemma~\ref{lem:measure},
  are not the same in both cases.  However, the $i$-th entry in the
  diagonal is given by
  \begin{align}
    \bra{i}\rho_0\ket{i}
    &= \bra{i}\:\left[\co(\ket{\Phi})\co(\bra{\Phi})-\!\we^*(\ket{\Phi})\we(\bra{\Phi})\right]\:\ket{i} \nonumber \\
    &= \bra{i}\co(\ket{\Phi})\:\co(\bra{\Phi})\ket{i}-\bra{i}\we^*(\ket{\Phi})\:\we(\bra{\Phi})\ket{i}\nonumber\\
    &= \co(\Phi_i)\co(\Phi_i^\star) - \we^*(\Phi_i)\we(\Phi_i^\star) \nonumber\\
    &= \co(\Phi_i)\co^*(\Phi_i) + \we^*(\Phi_i)\we(\Phi_i) \nonumber \\
    &= \modulus{\co(\Phi_i)}^2 + \modulus{\we(\Phi_i)}^2 = \modulus{\Phi_i}^2
\intertext{where $\Phi_i$ is the $i$-th coordinate of $\ket{\Phi}$, and we use 
the properties of $\co$ and $\we$ in Equation~\ref{eq:coweIdent}.  We also have, }
    \bra{i}\rho_1\ket{i}
    &= \bra{i}\:\left[-\we(\ket{\Phi})\we^*(\bra{\Phi})+\!\co^*(\ket{\Phi})\co^*(\bra{\Phi})\right]\:\ket{i} \nonumber \\
    &= -\bra{i}\we(\ket{\Phi})\:\we^*(\bra{\Phi})\ket{i}+\bra{i}\co^*(\ket{\Phi})\:\co^*(\bra{\Phi})\ket{i}\nonumber\\
    &= -\we(\Phi_i)\we^*(\Phi_i^\star) + \co^*(\Phi_i)\co^*(\Phi_i^\star) \nonumber\\
    &= \we(\Phi_i^\star)\we^*(\Phi_i^\star) + \co(\Phi_i^\star)\co^*(\Phi_i^\star) \nonumber \\
    &= \modulus{\we(\Phi_i^\star)}^2 + \modulus{\co(\Phi_i^\star)}^2 = \modulus{\Phi_i^\star}^2
     = \modulus{\Phi_i}^2
  \end{align}
\end{proof}

\section{Considerations and Consequences} 
\label{sec:Consider}

\subsection{Complexity of Simulation}

In terms of simulation resources, the situation is similar to that of
real computing.  Circuit width is increased by only one, but circuit
depth can be equal to the circuit size in the worst case.

For circuit size, however, we have to make a slight distinction.
While the number of $(d+1)$-ary gates in the new circuit will be the
same as the number of $d$-ary gates in the original circuit, one might
not be satisfied with this type of gate count complexity for the
quantum circuit, given that we do not know $d$ and that we have very
small universal gates for quantum circuits.  In general, if we suppose
that the original circuit given to us is constructed with some set of
universal gates, then the simulation will depend on $d$, the number of
quaterbits in the largest gate in the universal set.  In particular,
if $d>3$ we might require to decompose such a gate $Q_g$ into a set of
elementary 3-, 2- or 1-qubit gates, universal for quantum computing.

We can assume without loss of generality that we are given a full description of $Q_g$ in
terms of its $2^d\times 2^d$ quaternion matrix.  We can then use the
generic method for decomposing the matrix for the image quantum
operator $U_g=\qh(Q_g)$ into our set of elementary gates.  Since $U_g$
is a $2^{d+1}\times 2^{d+1}$ matrix this might require $O(2^d)$ time,
and furthermore up to $2^{d+1}$ elementary gates might be required to
decompose $Q_g$.

If a ``nice'' universal set is being used where $d$ is a small
constant, then this decomposition will occur in $O(1)$ time and will
produce $O(1)$ extra gates.  Hence, we have that the total gate count
is not \emph{exactly} $n$, but is still in $O(n)$.  The circuit depth
which could already be as large as $s$, could be increased further by
gate decomposition, but again, only by a constant factor.

\begin{mycomment}
  IF LOTS OF TIME: This is where it gets messy because we did not do
  our homework, i.e.\ find a goddam universal set....
\end{mycomment}


While we have not gone through the exercise of looking for a finite
universal set of elementary gates which would be computationally
universal for the symplectic group, we believe that one exists.  Even
without the luxury of a finite universal set, it would be in principle
sensible to define a computational model using quaternionic gates, as
long as the description of all circuits (and their gates) is of
limited size and can be uniformly generated.  In fact, our results do
not need the existence of a universal set; they just would make the
computing model more ``realistic.''

On the other hand, let us also consider a variety of quaternionic
circuits which includes gates of arbitrary degree ---since we cannot
show a ``nice'' universal set with constant degree gates, let us do so
for the sake of completeness.  In that case, if the circuit
description has size polynomial in $n$, then the description of $Q_g$
must also be of polynomial size, and this puts an upper bound on $d$,
i.e.~$d=O(\log n)$.  Thus, in the worst, case, we can have that each
$Q_g$ will require $2^{d+1}=O(n)$ elementary quantum gates, all in
series, with a resulting $O(n)$ depth and size overhead \emph{for
  each} gate.  Computing these decompositions would take time at most
$O(n)$ per gate.  We summarise these results in
Table~\ref{tab:complexity}.

\begin{table}[h]
  \centering
  \begin{tabular}{r|c|c}
        & Quaternionic circuit & Quantum circuit \\ \hline
  width & $n$ & $n+1$ \\
  size  & $s$ & $s2^{d+1}$ \\
  depth & $t$ & $t2^{d+1}$
  \end{tabular}
  \caption[Resources Needed to Simulate a Quaternionic Circuit]%
  {The overall resources needed to simulate a quaternionic circuit 
     built with $d$-ary gates, with a quantum circuit built with 2-ary gates.}
  \label{tab:complexity}
\end{table}

We stress the fact that this is a worst case scenario due to the fact
that we cannot bound $d$ by a constant, as we have not yet shown any
universal set of quaternionic gates.  If we did, then $d=O(1)$, and
the results would be the same, up to a constant, as those for
Theorem~\ref{thm:RebitSim}.







\subsection{Interpretations of the Quaternionic Model}

\begin{mycomment}
Explain effect of non-distributivity of tensor and matrix product:
Cannot ``disentangle'' even if separated to start with...
- Conjecture this is probably true for the kind of matrices that
we are dealing with??
\end{mycomment}

Because of the similarity of the constructions of Theorems
\ref{thm:RebitSim} and~\ref{thm:QuaterSim}, we can give similar
interpretations to the role of the extra required top qubit.  More
concretely, if we label the basis of the $2N$-dimensional complex
Hilbert space as $\ket{b_c}=\qh_0(\ket{b})$ and
$\ket{b_w}=\qh_1(\ket{b})$, and order them accordingly, we can give
the same semantics to the extra wire required by the simulation.
This is, the extra qubit is at the top of the simulating circuit, and
in a similar way as before keeps track of the ``phase'' information
between both orthogonal subspaces of the complex Hilbert space spanned
by the $\ket{b_c}$ and $\ket{b_w}$ base vectors.  In this case,
however, this information requires the full ``power'' of a qubit, and
not just a rebit.  This is due to the fact that the phase information
is defined by a unit quaternion, which cannot be represented by just
one angle (as is the case for a unit complex number).

We can infer, that with this same method it is not possible to
simulate an $n$ quaterbit circuit with only $n+1$ rebits.  The
following corollary, however, shows that just one extra rebit is
sufficient.

\begin{corollary}\label{cor:quaterReal}
  Any temporal chain $\sigma$ of an $n$-quaterbit quaternionic circuit
  can be exactly simulated by an $(n+2)$-rebit real circuit.
\end{corollary}

Two proofs are possible.  First, we can simply combine the results of
Theorems \ref{thm:RebitSim}~and \ref{thm:QuaterSim}.  More
interestingly, however, a direct proof is possible by using the
standard representation of quaternions as $4\times 4$ real matrices,
which suggests the following tensor~$\qrtensor$
\begin{equation}
  \qrtensor \eqdef \left(\begin{array}{rrrr}
       \re &  \im & -\km & -\jm \\
      -\im &  \re & -\jm &  \km \\
       \km &  \jm &  \re &  \im \\
       \jm & -\km & -\im &  \re
      \end{array}\right)
  \label{eq:q2tensor}
\end{equation}
where $\jm(\qa)\eqdef a_2$ and $\km(\qa)\eqdef a_3$ are the ``other''
imaginary parts of quaternion $\qa$.  This tensor induces a group
isomorphism from $\Sp(N)$ to $\SO(4N)$, which has all the properties
required for the simulation to be sound.  

\begin{figure}[h]
  \centering
  \subfigure[]{%
    \label{fig:non-commut1}
    \fbox{\includegraphics[scale=1]{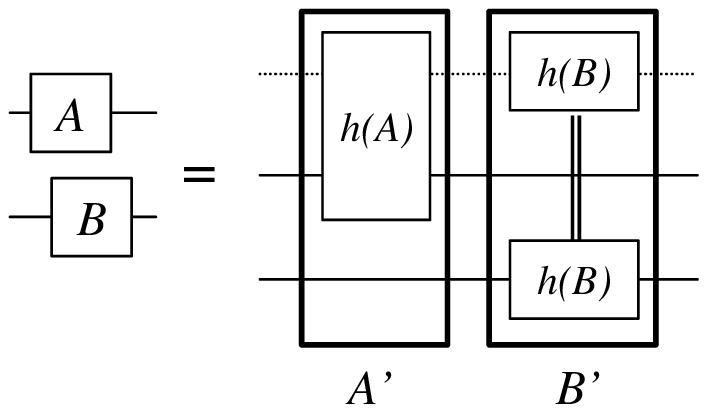}}
  }\hspace{1em}
  \subfigure[]{%
    \label{fig:non-commut2}
    \fbox{\includegraphics[scale=1]{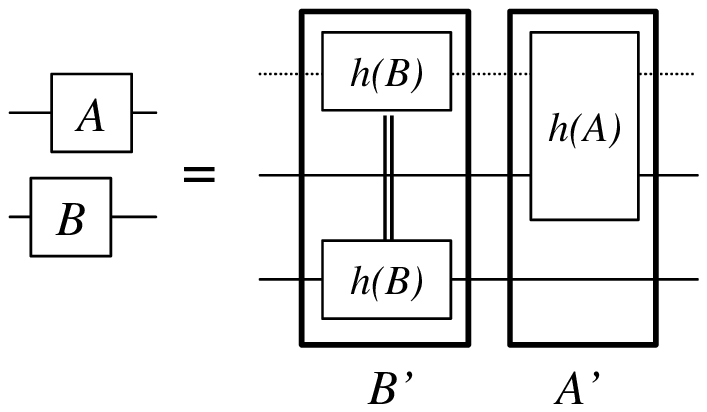}}
  }
  \caption{%
    More effects of quaternionic non-commutativity on quaternionic
    circuits.  In this simple 2-quaterbit example, the global circuit
    operator will depend on whether gate $A$ is executed
    \subref{fig:non-commut1} ``before'' or \subref{fig:non-commut2}
    ``after'' gate $B$.}
\label{fig:non-commut}
\end{figure}

Within the context of this simulation, the fact that the output of a
quaternionic circuits depends on the order of evaluation of the gates
becomes painfully obvious.  Consider a simple quaternionic circuit
with two 1-quaterbit gates $A$ and $B$ acting in parallel on two
separate quaterbits.  Let us consider a complex qubit simulation, as
illustrated in Figure~\ref{fig:non-commut}, where $A'$ and $B'$ are
the in-context unitary complex operators simulating them,
respectively.  Since we do not expect $A'$ and $B'$ to commute in
general, the simulation will produce two different results, depending
on which of the two gates is placed before.

However, since we have shown that the simulation is always accurate,
this ambiguity exists \emph{even if no simulation ever happens}.  In
other words, this non-local time dependence is a natural property of
quaternionic systems.  

From an Information Theory point of view, this non-local dependence
resembles entanglement, except that it somehow comes for ``free'':
\footnote{Not to be confused with \emph{free entanglement}, a term
  which is sometimes used to refer the opposite of \emph{bound
    entanglement}.} even if the initial state is completely
uncorrelated and unentangled (i.e. a product state), we can obtain a
global state in which this dependence exists even \emph{without
  performing any non-local operations}.  The same is not possible in
standard complex-based qubits Quantum Information Theory.  Even though
we dare not call this curiosity ``entanglement'', the fact is that it
has important consequences on a quaternionic Information Theory.  In
particular, it would allow Alice and Bob to perform a reasonable form
of unconditionally secure bit commitment.  In our example, let Alice
and Bob hold two completely unentangled and uncorrelated quaterbits.
At commitment time, Bob announces that he is about to do operation $B$
on his quaterbit.  Alice then commits to 0 or 1 by performing $A$ on
her quaterbit before or after Bob does $B$, respectively.  The
commitment is opened by Alice and Bob putting their quaterbits
together and measuring them and verifying which of the two possible
outcomes came out.  Of course, Bob cannot open before Alice provides
her quaterbit and Alice cannot change her mind after Bob has done $B$
without it being detected by Bob at the opening stage.  Again, this is
not possible in the standard Quantum Information Theory
\cite{May97,LC98}.

According with current attempts to reformulate Quantum Mechanics
purely in terms of Information Theory (\cite{Fuc02,Har01}, among
others), the possibility of bit commitment alone would be sufficient
to rule out quaternionic models as reflecting physical reality.
However, even if this programme does not succeed, one might argue that
the non-local time-dependence exhibited by quaternionic models is in
itself non-physical enough to rule them out, as they can be seen as a
mild violation of the usual causality properties of Nature.

\section{Final Conclusions and Further Questions}
\label{sec:Final}


\subsubsection*{The possibility of a quaternionic speed-up?}

We have shown how a somewhat sensible model of computing can be
constructed using quaternionic amplitudes.  A crucial characteristic
of this model is that due to the non-commutativity of quaternions, the
output to the circuit will depend on the ``evaluation path'' of the
circuit, as there is no unique circuit operator for all possible ways
of re-combining the gates.  However, any such ordering of gates
generates a well defined output, which we have shown can be simulated
exactly and efficiently by a quantum circuit of similar size and
width.  This was our main result, which was inspired on a new proof we
constructed for the equivalence of complex and real circuits.  Despite
this somewhat strange and unexpected parametrised definition of
quaternionic computing, what this result in essence tells us is that
all of these paths along the space-time continuum somehow have the
same computational power, and furthermore that they can be
independently simulated in an efficient manner by a standard quantum
computer.


We can interpret Theorem~\ref{thm:QuaterSim} as a general result on
quaternionic physical models as follows.  If somehow Nature chooses
and prefers \emph{one} of the possible paths of evolution through the
state space, then Nature's behaviour on such quaternionic systems can
be efficiently simulated by a quantum system of similar complexity.
This, provided that we somehow know which path is preferred.  If this
were indeed the case (for example, because the physicists would tell
us so), we complexity theorists might rub our hands together in
satisfaction and further sing to the robustness of the \BQP\
complexity class.

But what if Nature somehow \emph{did not} prefer nor chose one these
paths, but somehow \emph{followed them all at the same time}.  Would
there be any mechanism by which the results of the different
computations would be weighed (by probabilities or probability
amplitudes)?  Could these paths interfere with each other, in a
similar fashion as in quantum models?  Destructively?  And if it were
the case, could we ultimately harness such extra parallelism to
achieve a speed-up beyond those achievable with quantum computation
models?

\subsubsection*{Ruling out quaternions and real numbers}

The skeptics and realists among the readership might argue that all of
these questions are completely sterile and void of interest.  Despite
the fact that a quaternionic version of Quantum Mechanics has been
proposed \cite{Adl95}, we do not really know where or how Nature would
exhibit such ``quaternionic behaviour'' and even less how to harness
it.  If our only objective was to one day build a quaternionic
computer, we the authors would be the first to agree with such
skeptics.  Nevertheless, we believe that one of the major
contributions of this work has been to find and identify a simple and
easily explainable potential reason why there \emph{should not be}
quaternion amplitudes involved in Nature: the asymmetry of the
possible evolution paths between two space-time events, even without
relativistic effects.  This, the physicists might argue, is the
violation of some fundamental principle, and hence not possible or
likely.  From an information theoretic point of view, the possibility
of bit commitment provide reason enough to rule them out as a
``natural'' model.  These realisations are also in line with work of
Lucien Hardy \cite{Har04}, that also displays the unnaturality of
quaternionic or real models by considering the number of degrees of
freedom involved in the composition of such sub-systems.

From a purely information-theoretic point of view, there was already
some work including that of Caves, Fuchs and Schack~\cite{CFS02} and
Vlasov~\cite{Vla01} identifying some non-trivial differences between
standard and real-number or quaternion-based Information Theory.  In
the context of this paper, it is interesting to note that the
converses of Theorems~\ref{thm:RebitSim}~and \ref{thm:QuaterSim} are
not necessarily true.  Not all $(n+1)$-rebit/qubit circuits can be
simulated by $n$ qubit/quaterbit circuits, which stems from the fact
that $h$ and $\qh$ do not span the whole $\SO(2N)$ and $\Sp(2N)$,
respectively, as a simple counting argument shows.  While from a
complexity point of view requiring one extra qubit/rebit is not a big
deal, this asymmetry between the models might make a difference in
other quantum information processing tasks.  For example, one might
ask how many classical bits are required to teleport a quaterbit,
whether using quaterbits affects the various quantum channel capacity
measures, how communication complexity is affected, etc.  Furthermore,
while we have concentrated here our discussion on departures of the
quaternionic models from the standard one, the same fundamental
questions as above can be asked of rebits.  We believe that a study
and discussion of the real number case would also be interesting and
shed even more light on the physicality (or lack thereof) of such
models, quaternionic or real.

\subsubsection*{Completing the Algebraic ``Big Picture''}

Finally, we believe that the continued study of non-standard algebraic
models such as these, based for example on the octonions or even
possibly finite fields will also bear fruits in that direction.  More
concretely, we hope that, at the very least, we might be able to
provide more examples of ``weird properties'' which might discount
these models as ``unnatural'', doing so more easily in the language of
Information Theory than in that of Physics.

From a Complexity Theory point of view, this would also be of value in
further completing the algebraic ``big picture'' of complexity classes
painted in \cite{BFH02,BFH04}.  This picture, so far, gives evidence
of how little the actual amplitude structure does to change
computational power, and further points to what we believe is the
ultimate cause for the ``quantum speedup'', the possibility for
probability amplitudes to destructively interfere.

\subsection*{Acknowledgements}
The authors wish to thank David Poulin, Gilles Brassard and Michel
Arsenault for fruitful discussions on this topic and particularly
Lucien Hardy for pointing out other ``strange'' information-theoretic
properties of quaternions.  The authors would also like to thank
Lucien Hardy, Chris Fuchs, Scott Aaronson, and Alexander Vlasov for
pointing further literature on quaternion-based information theory and
quantum mechanics.

\bibliographystyle{plain}
\bibliography{strings,qcomp,books,qinfotheory}

\end{document}